%% file: vilkov.tex
\documentclass[vecphys]{svmult}
\usepackage{latexsym}
\newcommand{\f}{\varphi}
\makeatletter
\@addtoreset{equation}{section}
\makeatother
\begin{document}
\title*{Expectation Values and Vacuum Currents\\
of Quantum Fields\thanks{The course of 4 lectures given
at Coll\`ege de France in May 2006.}}
\titlerunning{Expectation Values and Vacuum Currents}
\author{G.A. Vilkovisky}
\authorrunning{Vilkovisky} 
\institute{Lebedev Physical Institute, Leninsky Prospect 53,
Moscow 119991, Russia.
\texttt{vilkov@lebedev.ru}}
\maketitle
\begin{abstract}
Theory of expectation values is presented as an alternative to 
S-matrix theory for quantum fields. This change of emphasis is
conditioned by a transition from the accelerator physics to
astrophysics and cosmology. The issues discussed are the time-loop
formalism, the Schwinger--Keldysh diagrams, the effective action,
the vacuum currents, and the effect of particle creation.
\end{abstract}
\section*{Introduction}
High-energy physics will probably have to undergo major changes.
The accelerators will cease being its experimental base, and it
will become a part of astrophysics. Simultaneously, the S-matrix
will cease being the central object of high-energy theory because
the emphasis on this object is entirely owing to the accelerator
setting of the problem. If there is a background radiation that
originates from some initial state in the past, then where is
the S-matrix here? Astrophysics and cosmology offer the
evolution problems rather than the scattering problems. The
gravitational collapse is a typical initial-value problem.
It is such by its physical setting irrespective of whether
the state of the system is classical or quantum. The nature
of measurement also changes. No final state is prepared.
One measures observables like temperatures or mechanical
deflections and subjects these measurements to a statistical
treatment to obtain the value of the observable. This means
that one measures expectation values in the given
initial state.
S-matrix theory should give way to expectation-value theory.

There is a proof that accelerator physics is dead: Gabriele
Veneziano is leaving CERN for Coll\`ege de France. At this
historic moment, my mission is to convert him into a new
faith. The present preaching consists of 4 lectures:
\begin{enumerate}
\item Formal aspects of expectation-value theory.
\item The in-vacuum state and Schwinger--Keldysh diagrams.
\item The effective action.
\item Vacuum currents and the effect of particle creation.
\end{enumerate}
Literature to Lectures 1 and 2 is in \cite{1}--\cite{16}.
Additional literature to Lecture~3 is in \cite{17}--\cite{41}
and to Lecture 4 in \cite{42}--\cite{56}.
\input lecture1.tex

\input lecture2.tex
\input lecture3.tex
\input lecture4.tex

\end{document}

%% file: lecture1.tex
\section[Lecture 1]{Formal Aspects of Expectation-Value Theory}
\label{sec:1}
{\renewcommand{\theequation}{1.\arabic{equation}}
\subsubsection{Vocabulary}
In these lectures,
\begin{equation}
{\hat\varphi}^i
\end{equation}
denotes the quantum field. It is an operator function on a 
given differentiable manifold (referred to below as the base manifold), 
and $i$ is a point of this manifold.
Generally, ${\hat\f}^i$ is a collection of fields, and then $i$
is a set containing also the indices labelling these fields. 
The hat designates an operator.
The ${\hat\f}^i$ is an operator in a Hilbert
space which is not granted. The workers have to build it 
with their own hands
as a representation of the algebra of ${\hat\f}$'s. For simplicity,
${\hat\f}^i$ will be assumed boson and real (self-adjoint) but
otherwise arbitrary. 

The starting point is an operator equation 
for ${\hat\f}^i$ 
\begin{equation}
S_i({\hat\f})+J_i=0
\end{equation}
which is understood as an expansion. It is meant that there is
a c-number function $S_i(\f)$ understood as a collection of its
Taylor coefficients at some c-number point of configuration space:
\begin{equation}
S_i(\f)=\sum_{n=0}^{\infty}\frac{1}{n!}
S_{ij_1 \cdots j_n}(c)(\f-c)^{j_1}\ldots(\f-c)^{j_n}\;,
\end{equation}
and one replaces $\f^j$ in this expansion with an operator.
Which c-number field $c^j$ will be used for this expansion
does not matter because it will always sum with the operator
$({\hat\f}-c)^j$ to make the full quantum field. The expansion point
$c^j$ is often called "background field", and there
has been much emphasis on it. In fact it is completely
immaterial. I shall never make this expansion 
explicitly but I shall keep explicit the c-number term of the
equation: a source $J_i$.

Important are only the following three points.
\begin{itemize}
\item[(1)] The function $S_i(\f)$ is local, i.e., it depends
only on $\f$ and its finite-order derivatives at the point $i$.
\item[(2)] The function $S_i(\f)$ is a gradient:
\begin{equation}
S_i(\f)=\frac{\delta}{\delta\f^i}S(\f)\;,
\end{equation}
i.e., there exists an action $S(\f)$ generating the 
operator field equations.
For its derivatives the following notation will be used:
\begin{equation}
S_{i_1\cdots i_n}(\f)=\frac{\delta}{\delta\f^{i_1}}\cdots
\frac{\delta}{\delta\f^{i_n}}S(\f)\;.
\end{equation}
Of course, only the total action matters:
\begin{equation}
S_{\mbox{\scriptsize tot}}=S(\f)+\f^iJ_i\;.
\end{equation}
\item[(3)] There is a special condition on the matrix of second
derivatives of $S(\f)$. I shall refer to this continuous
matrix as $S_2$:
\begin{equation}
S_{ij}(\f)\equiv S_2(\f)\;.
\end{equation}
By locality, $S_2$ is the kernel of some differential operator
on the base manifold for which I shall use the same notation $S_2$.
It is required that $S_2$ admit a well-posed Cauchy problem
in which case it has the unique advanced and retarded inverses
(Green's functions) $G^+$ and $G^-$:
\begin{equation}
S_{ij}G^{\pm jk}=-\delta^k_i\;,\qquad G^{+jk}=G^{-kj}\;.
\end{equation}
Because $S_2$ is symmetric, the advanced inverse is the transpose
of retarded.
\end{itemize}

One may think of $S_2$ as of a second-order hyperbolic operator
which it will in fact be below but the scheme is more general.
It is formalism-insensitive. One's field equations may have
the second-order differential form or the first-order
differential form, -- the scheme will work anyway. The importance
of the operator $S_2$ is in the fact that it determines the
linear term of the field equations and, therefore, governs
the iteration procedures. Commute ${\hat\f}^i$ with the
field equations. Obtained will be a linear homogeneous
equation for the commutator $[{\hat\f}^i,{\hat\f}^j]$.
Consider the respective inhomogeneous equation and its
two iterative solutions: one with the advanced inverse
for $S_2$ and the other one with retarded. The equation for the
commutator is solved by their difference:
\begin{equation}
[{\hat\f}^i,{\hat\f}^j]={\rm i}\hbar\left(G^{+ij}(c)
-G^{-ij}(c)\right)+O({\hat\f}-c)\;.
\end{equation}
In this way the algebra of ${\hat\f}$'s is built as an operator
expansion. This is the quantization postulate.

By the setting of its Cauchy problem, the operator $S_2$ introduces
the concept of causality. If $S_2$ is a second-order hyperbolic
operator, this is the usual relativistic causality. But in any
case the base manifold will be foliated with the Cauchy surfaces
of the operator $S_2$. They will be denoted as $\Sigma$.

A function of ${\hat\f}$ that involves ${\hat\f}$ on only one
Cauchy surface 
\begin{equation}
Q({\hat\f})=Q({\hat\f}\Bigl|_\Sigma)
\end{equation}
will be called local observable. A state defined as an eigenstate
of local observables
\begin{equation}
Q({\hat\f}\Bigl|_\Sigma)|\;\;\rangle=q|\;\;\rangle
\end{equation}
will be called local state. This latter name may be confusing
because the state is, of course, a global concept, and I am using
the Heisenberg picture. But the local state is {\it associated}
with a given $\Sigma$:
\begin{equation}
|\;\;\rangle=|\Sigma,q\rangle\;.
\end{equation}
Of course, for it to be defined, one needs a complete set of
commuting local observables. I call the $Q$'s observables but
they may not even be Hermitian. And I shall consider them linear 
in ${\hat\f}$. If they are nonlinear, I shall make a local
reparametrization of the field variables so as to make them linear.

In fact, if one has a complete set of commuting local observables,
one has already built a Hilbert space. A linear combination
\begin{equation}
|\Sigma\rangle=\int dq\,\Psi(q)|\Sigma,q\rangle
\end{equation}
is also a local state associated with $\Sigma$ provided that
the function $\Psi(q)$ is external, i.e., independent of the 
quantum field ${\hat\f}^i$.

Our goal is to learn how to calculate expectation values of
field observables in a local state, and I shall concentrate
on the expectation value
\begin{equation}
\langle\Sigma|{\hat\f}^i|\Sigma\rangle\;.
\end{equation}
However, we shall save the effort if we consider another
problem first. Namely, let us recall what would we do in the case
of two local states associated with different Cauchy surfaces:
\begin{equation}
|\Sigma_1,q_1\rangle=|1\rangle\;,\qquad
|\Sigma_2,q_2\rangle=|2\rangle\;,
\end{equation}
$$
\Sigma_2>\Sigma_1\;.
$$
Here and below, "greater" is a notation for "later".
\subsubsection{The Quantum Boundary-Value Problem}
In the problem where given are two local states (1.15), 
the field's expectation value is replaced
with the scalar product
\begin{equation}
\frac{\langle2|{\hat\f}|1\rangle}{\langle2|1\rangle}
\stackrel{\mbox{def}}{=}\langle\f\rangle
\end{equation}
which I shall call mean field although it is not mean in any state.

If our goal was the scalar product (1.16), we would use the Schwinger
principle
\begin{equation}
\delta\langle2|1\rangle={\rm i}\langle2|
\delta S_{\mbox{\scriptsize tot}}|1\rangle\mbox{ or zero}
\end{equation}
whose meaning is this. Consider a variation in the Taylor coefficients
of the field equations, i.e., in the functional form of the total
action. The solution for ${\hat\f}^i$ will respond and will induce
a change in the functions $Q({\hat\f})$ which will induce a change
in their eigenstates, and finally there will be a change in the
amplitude $\langle2|1\rangle$ induced by a change in the action.
The Taylor coefficients are local. They can be varied in the
region between $\Sigma_1$ and $\Sigma_2$ or outside this region.
The Schwinger principle (1.17) says that, if they are varied
outside, the variation of the amplitude is zero. Otherwise, this
variation is expressed through the variation of the action
by (1.17).

The Schwinger principle is a consequence of the commutation
relations but it can also be taken for the first principle
because one does not need anything else. For many purposes
(but not all) it suffices to use a specific case of (1.17): 
a freedom of varying the source $J$. The result of this use is
\begin{equation}
\frac{\delta}{\delta{\rm i}J_{j_1}}
\cdots
\frac{\delta}{\delta{\rm i}J_{j_n}}
\langle2|1\rangle=\left\{
\begin{array}{l}
\langle2|\overleftarrow{T}\left({\hat\f}^{j_1}\ldots
{\hat\f}^{j_n}\right)|1\rangle,
\mbox{ if }\Sigma_2>j_1,\ldots j_n>\Sigma_1\;,\\
0,\mbox{ otherwise}\;.\\
\end{array}
\right.
\end{equation}
Here $T$ orders the operators ${\hat\f}^k$, $k\in\Sigma_k$,
chronologically, i.e., places them in the order of following 
of their $\Sigma_k$, and the arrow
over $T$ points the direction of growth of the time $\Sigma$.

Let us come back to the operator field equations. Since all
${\hat\f}$'s in these equations are at the same point,
one can formally insert in (1.2) the sign of chronological
ordering:
\begin{equation}
\overleftarrow{T}S_i({\hat\f})+J_i=0\;.
\end{equation}
One may worry about additional terms in (1.19) stemming from
the distinction between the chronological and ordinary operator products,
and the noncommutativity of $\overleftarrow{T}$ with the derivatives
in the Taylor coefficients of the equations. Because the operators
in the products are at the same point, these terms are
ambiguous expressions whose handling depends on
the formalisms and procedures used. There is always a happy end:
these terms cancel and help to cancel similar terms appearing
in the subsequent calculations. Therefore, it makes sense to use
such formalisms and procedures that these terms do not appear
at all. This is the approach that I shall follow.

Sandwiching the equation (1.19) between the states $\langle2|$ and
$|1\rangle$, and using (1.18), one obtains the following equation
for the amplitude:
\begin{equation}
\left(S_i\left(\frac{\delta}{\delta{\rm i}J}\right)
+J_i\right)\langle2|1\rangle=0\;.
\end{equation}
Multiply it from the left with $\langle2|1\rangle^{-1}$ and pull
the factors $\langle2|1\rangle$ in the argument of $S_i$ using
the fact that this is a unitary transformation:
\begin{equation}
\left(S_i\left(
\langle2|1\rangle^{-1}
\frac{\delta}{\delta{\rm i}J}
\langle2|1\rangle\right)
+J_i\right)1=0\;.
\end{equation}
In the argument, commute the operators:
\begin{equation}
\left(S_i\left(
\frac{\delta\ln\langle2|1\rangle}{\delta{\rm i}J}+
\frac{\delta}{\delta{\rm i}J}\right)
+J_i\right)1=0
\end{equation}
and use that by (1.18)
\begin{equation}
\frac{\delta\ln\langle2|1\rangle}{\delta{\rm i}J_k}=
\langle\f^k\rangle\;.
\end{equation}
The result is the following equation for the mean field:
\begin{equation}
\left(S_i\left(
\langle\f\rangle+
\frac{\delta}{\delta{\rm i}J}\right)
+J_i\right)1=0\;.
\end{equation}

Equation (1.24) differs from the classical field equation by the
operator addition 
$\delta/\delta{\rm i}J$
to $\langle\f\rangle$. When this operator
addition acts on $1$, its effect is zero, but it will act also on
$\langle\f\rangle$ because the summands 
$\langle\f\rangle$ and
$\delta/\delta{\rm i}J$
do not commute. Where in (1.24) is the Planck constant? It is easy
to see by dimension that $\hbar$ is just in front of
$\delta/\delta{\rm i}J$. Therefore, if one wants to expand the
equations in $\hbar$, one should expand them in
$\delta/\delta{\rm i}J$.

The problem boils down to expanding a function $f(A+B)$ in $B$
when $A$ and $B$ do not commute. It suffices to expand the
exponential function since one can write
\begin{equation}
f(A+B)=f\left(\frac{d}{dx}\right)\left.{\rm e}^{(A+B)x}\right|_{x=0}
\end{equation}
or, equivalently,
\begin{equation}
f(A+B)=\left.{\rm e}^{(A+B)d/dx}f(x)\right|_{x=0}\;.
\end{equation}
For the exponential function one has the identity
\begin{equation}
{\rm e}^{(A+B)x}={\rm e}^{Ax}\left(1+\int\limits_0^x dy\,
{\rm e}^{-Ay}B{\rm e}^{(A+B)y}\right)
\end{equation}
which makes the expansion possible. This all works well if
the series of commutators
\begin{equation}
{\rm e}^{-A}B{\rm e}^A= B+[B,A]+\frac{1}{2!}[[B,A],A]
+\frac{1}{3!}[[[B,A],A],A]+\cdots
\end{equation}
terminates somewhere as in our case. Indeed, if
$\langle\f\rangle=A$ and
$\delta/\delta{\rm i}J=B$, then
\begin{equation}
[[B,A],A]=0\;.
\end{equation}
Under condition (1.29) one obtains for an arbitrary function:
\begin{equation}
f(A+B)=f(A)+f'(A)B+\frac{1}{2}f''(A)[B,A]+O(B^2)\;.
\end{equation}
As compared to the ordinary Taylor expansion, there are
several additional terms with commutators at each order.

A use of the result above in equation (1.24) gives
\begin{equation}
S_i(\langle\f\rangle)+\frac{1}{2}S_{ijk}(\langle\f\rangle)
\frac{\delta\langle\f^j\rangle}{\delta{\rm i}J_k}
+O(\hbar^2)=-J_i\;,
\end{equation}
\begin{equation}
S_{ij}(\langle\f\rangle)
\frac{\delta\langle\f^j\rangle}{\delta J_k}
=-\delta^k_i+O(\hbar)\;.
\end{equation}
Here the second equation is obtained by differentiating the
first one, and it tells us what is
$\delta\langle\f\rangle/\delta J$. Up to $O(\hbar)$, it
is some Green's function of the operator $S_2$. Denote this
Green's function as
\begin{equation}
\frac{\delta\langle\f^j\rangle}{\delta J_k}
=G^{jk}+O(\hbar)\;.
\end{equation}

One can work to any order but I shall stop here. {\it We obtain
closed equations for the mean field}:
\begin{equation}
S_i(\langle\f\rangle)+\frac{1}{2{\rm i}}S_{ijk}(\langle\f\rangle)
G^{jk}(\langle\f\rangle)
+O(\hbar^2)=-J_i\;,
\end{equation}
\begin{equation}
S_{ij}(\langle\f\rangle)
G^{jk}(\langle\f\rangle)
=-\delta^k_i\;.
\end{equation}
The second term in (1.34) is the loop
\begin{equation}
S_i(\langle\varphi\rangle)+{}
\parbox{30pt}{
\begin{picture}(30,20)
\thicklines
\put(20,10){\circle{20}}
\put(0,10){\line(1,0){10}}
\put(0,12){$\scriptstyle i$}
\end{picture}
}
{}+O(\hbar^2)=-J_i\;,
\end{equation}
all elements of the loop being functions of
$\langle\f\rangle$. But two questions remain to be answered:
\begin{itemize}
\item[(i)] Which Green's function is $G$?
\item[(ii)] What are the boundary conditions to the mean-field
equations?
\end{itemize}

The answers are again in the Schwinger principle. Equation (1.18)
tells us what are $G$ and $\langle\f\rangle$:
\begin{equation}
\frac{1}{{\rm i}}G^{jk}=\frac{
\langle2|\overleftarrow{T}\left({\hat\f}^j
{\hat\f}^k\right)|1\rangle}{\langle2|1\rangle}
-\langle\f^j\rangle\langle\f^k\rangle+O(\hbar)\;,
\end{equation}
\begin{equation}
\langle\f^j\rangle=
\frac{\langle2|{\hat\f}^j|1\rangle}{\langle2|1\rangle}\;.
\end{equation}
Multiply these expressions by the coefficients that make
the linear $Q$ out of $\f$:
\begin{equation}
Q({\hat\f})=k_j{\hat\f}^j\;,
\end{equation}
and send $j$ either to $\Sigma_1$ or to $\Sigma_2$.
By the definition of the states $|1\rangle$ and $|2\rangle$,
one obtains
\begin{equation}
Q(\langle\f\rangle\Bigl|_{\Sigma_1})=q_1\;,\qquad
Q(\langle\f\rangle\Bigl|_{\Sigma_2})=q_2\;,
\end{equation}
\begin{equation}
k_jG^{jk}\Bigl|_{j\in\Sigma_1}=0\;,\qquad
k_jG^{jk}\Bigl|_{j\in\Sigma_2}=0\;.
\end{equation}
From (1.37) it follows also that
\begin{equation}
G^{jk}=G^{kj}\;.
\end{equation}
The Green's function $G$ is symmetric and completely determined
by the boundary conditions (1.41). This completes the determination
of the mean-field equations (1.34), and for these equations one
arrives at a boundary-value problem with the boundary conditions
(1.40). As a result, the quantum boundary-value problem is reduced
to a c-number boundary-value problem. I say "c-number" rather than
"classical" because there are differences, and one is the presence
of terms $O(\hbar)$ in the equations, but, as far as the setting
of the problem is concerned, there is no difference. One arrives
at the same boundary-value problem for the observable field as
in the case of the classical states.

Note that the Green's function $G$ and, thereby, the mean-field
equations do not depend on the eigenvalues $q$. The eigenvalues
appear only in the boundary conditions to the equations.
However, $G$ depends on the choice of the observables $Q$
themselves and, through them, on the choice of the states
$|1\rangle$ and $|2\rangle$. Therefore, the mean-field equations
are state-dependent.

Although the Green's function $G$ depends on the choice
of the states, it possesses
two universal properties. One has already been mentioned: $G$
is always symmetric. The other one is this. Let us make a variation
in the operator $S_2$ and find out how does $G$ respond:
$$
S_2G=-1\;,
$$
$$
S_2\delta G=-\delta S_2G\;,
$$
$$
\delta G=?
$$
To answer this question, one can use the Schwinger principle again.
The result is the following {\it variational law}:
\begin{equation}
\delta G=G\delta S_2G\;,
\end{equation}
and this law is universal. It is the same for all 
boundary-value problems.

The variational law (1.43) is remarkable. It is characteristic of 
finite-dimensional matrices. If a matrix has a unique inverse, then the
inverse obeys this law. This law is valid, for example, for
the inverse of an elliptic operator, i.e., for the Euclidean
Green's function. It is valid also for the advanced and
retarded Green's functions:
\begin{equation}
\delta G^+=G^+\delta S_2G^+\;,\qquad
\delta G^-=G^-\delta S_2G^-\;.
\end{equation}
But it is not valid generally, and, in the case of $S_2$, it
is exceptional.

The variational law for $G$ has an important implication.
Namely, let us differentiate the left-hand side of the
mean-field equations
\begin{equation}
\Gamma_i(\f)\equiv S_i(\f)+
\frac{1}{2{\rm i}}
S_{imn}(\f)G^{mn}(\f)+O(\hbar^2)
\end{equation}
to see if the result is symmetric. One obtains
\begin{eqnarray}
\frac{\delta\Gamma_i(\f)}{\delta\f^j}-
\frac{\delta\Gamma_j(\f)}{\delta\f^i}&\!=\!&
\frac{1}{2{\rm i}}
S_{imn}G^{m{\bar m}}G^{n{\bar n}}S_{{\bar m}{\bar n}j}
-(i\leftrightarrow j)+O(\hbar^2)\nonumber\\
&\!=\!&0+O(\hbar^2)\;.
\end{eqnarray}
This means that $\Gamma_i(\f)$ is a gradient, i.e., there exists
an action generating the mean-field equations:
\begin{equation}
\Gamma_i(\f)=\frac{\delta\Gamma(\f)}{\delta\f^i}\;.
\end{equation}
There is another way to arrive at the same conclusion.
Consider a function of the mean field defined by the
Legendre transformation
\begin{equation}
\Gamma(\langle\f\rangle)=
\frac{1}{{\rm i}}\ln\langle2|1\rangle-
\langle\f^k\rangle J_k
\end{equation}
where $J$ is to be expressed through $\langle\f\rangle$ by solving
equation (1.23). It is easy to see that this function satisfies
the equation
\begin{equation}
\frac{\delta\Gamma(\langle\f\rangle)}{\delta\langle\f^i\rangle}
=-J_i\;,
\end{equation}
and, therefore, its gradient is the left-hand side of the
mean-field equations.

$\Gamma(\f)$ is the effective action. Up to $\hbar^2$ it is
of the form
\begin{equation}
\Gamma(\f)=S(\f)+
\frac{1}{2{\rm i}}\ln\det G(\f)+O(\hbar^2)
\end{equation}
where the second term is the loop without external lines:
\begin{equation}
\Gamma(\varphi)=S(\varphi)+{}
\parbox{20pt}{
\begin{picture}(20,20)
\thicklines
\put(10,10){\circle{20}}
\end{picture}
}
{}+O(\hbar^2)\;.
\end{equation}
The effective action exists for any boundary-value problem 
but these actions are different for different such
problems. Only in the classical approximation, the action and the
equations are independent of the boundary conditions.

Let us go over to expectation values.
\subsubsection{The Quantum Initial-Value Problem}
In this problem, given is only one local state (which I shall
assume normalized). Since
the field operators are now sandwiched between the states
associated with one and the same $\Sigma$:
\begin{equation}
\langle1|(\cdots)|1\rangle\;,\qquad \langle1|1\rangle=1
\end{equation}
one cannot apply the Schwinger principle: there is no room
for varying the source. One can create this room artificially
by inserting a complete set of states associated with some
later $\Sigma$:
\begin{equation}
\langle1|1\rangle=\sum_q
\langle1|2q\rangle
\langle2q|1\rangle\;,
\end{equation}
$$
\Sigma_2>\Sigma_1
$$
but this alone will not help because the source is varied 
in both amplitudes, and these variations cancel. It will
help only if the two amplitudes in (1.53) are functions of
different sources, i.e., if, instead of (1.53), one introduces
a function of two independent sources, $J$ and $J^*$:
\begin{equation}
Z(J^*,J)=\sum_q
\langle1|2q\rangle_{J^*}
\langle2q|1\rangle_J\;.
\end{equation}
This amounts to considering two copies of the quantum field:
one with the source $J$, the other one with the source $J^*$,
and using in (1.54) the amplitudes of both. Then one can vary
only one source and, after that, make the sources coincident.
Using the Schwinger principle, one obtains
\begin{equation}
\left.\frac{\delta^nZ(J^*,J)}{\delta{\rm i}J_{j_1}
\cdots
\delta{\rm i}J_{j_n}}\right|_{J^*=J}=
\langle1|\overleftarrow{T}\left({\hat\f}^{j_1}\ldots
{\hat\f}^{j_n}\right)|1\rangle\;.
\end{equation}
In this way the expectation values can be calculated.

The technique of two sources is called time-loop formalism
because in expression (1.54) one goes forward in time, from
$\Sigma_1$ to some $\Sigma_2$, 
and then back from $\Sigma_2$ to $\Sigma_1$ but with another copy
of the quantum field.

For every partial amplitude in (1.54) we have equation (1.20)
\begin{equation}
\left(S_i\left(\frac{\delta}{\delta{\rm i}J}\right)
+J_i\right)\langle2q|1\rangle_J=0\;.
\end{equation}
Since the other amplitude in (1.54) does not depend on $J$,
we can linearly combine equations (1.56) to obtain
\begin{equation}
\left(S_i\left(\frac{\delta}{\delta{\rm i}J}\right)
+J_i\right)Z(J^*,J)=0\;.
\end{equation}
Only one source is active in this differential equation.
The other one is a parameter. Therefore, we can just
repeat the consideration above with $Z(J^*,J)$ in place of
$\langle2|1\rangle$, and in this way derive the mean-field
equations. We obtain the loop expansion of exactly the same
form as before:
\begin{equation}
S_i(\langle\f\rangle)+\frac{1}{2{\rm i}}S_{ijk}(\langle\f\rangle)
G^{jk}(\langle\f\rangle)
+O(\hbar^2)=-J_i\;,
\end{equation}
\begin{equation}
S_{ij}(\langle\f\rangle)
G^{jk}(\langle\f\rangle)
=-\delta^k_i\;,
\end{equation}
and in these loops we must make the sources coincident. There are
only two elements in all loops, $\langle\f\rangle$ and $G$.
Upon setting $J^*=J$, $\langle\f\rangle$ becomes the
genuine expectation value 
\begin{equation}
\langle\f^k\rangle=
\left.\frac{\delta\ln Z(J^*,J)}{\delta{\rm i}J_k}
\right|_{J^*=J}=
\langle1|{\hat\f}^k|1\rangle\;,
\end{equation}
and the matrix $G$ is given by the expression
\begin{equation}
\frac{1}{{\rm i}}G^{jk}+O(\hbar)=
\left.\frac{\delta^2\ln Z(J^*,J)}{\delta{\rm i}J_j
\delta{\rm i}J_k}\right|_{J^*=J}=
\langle1|\overleftarrow{T}\left({\hat\f}^j
{\hat\f}^k\right)|1\rangle
-\langle\f^j\rangle\langle\f^k\rangle\;.
\end{equation}
I am using for it the same letter $G$ but it is now
a different Green's function of the operator $S_2$.
Equations (1.58) with this Green's function in all loops
are the expectation-value equations.

The solution of the expectation-value equations is specified
completely by the initial conditions on $\Sigma_1$ following
from (1.60) but it is not easy to write these conditions 
down in the general terms. Only half of them is obvious:
the $Q$'s on $\Sigma_1$ are given. To obtain the other half,
one would need to find the variables canonically conjugate
to $Q$'s and calculate their expectation values 
on $\Sigma_1$.\footnote{Let $Q$'s be Hermitian, and let $P$'s
have c-number commutators with $Q$'s: ${[P,Q]={\rm i}}$. Then
the expectation values in the state (1.13) satisfy the initial
conditions
$$
\langle Q\Bigl|_\Sigma\rangle=\int dq\,
\overline{\Psi}(q)q\Psi(q)\;,
\qquad
\langle P\Bigl|_\Sigma\rangle={\rm i}\int dq\,
\overline{\Psi}(q)
\frac{\partial}{\partial q}\Psi(q)
$$
where the overline means complex conjugation. If both $Q({\hat\f})$
and $P({\hat\f})$ are linear, these are initial conditions
directly for $\langle\f\rangle$.} The same concerns 
the specification of the Green's function
$G$. This issue will be considered in the next lecture
where a different approach to it will be used.

Let us consider the state-independent properties of $G$.
First, as seen from (1.61), $G$ is symmetric for any initial-value
problem:
\begin{equation}
G^{jk}=G^{kj}\;.
\end{equation}
Second, one can apply the Schwinger principle to derive the 
variational law for $G$. At this point, the initial-value problem
differs significantly from the boundary-value problem. When the
operator $S_2$ is varied in the generating function (1.54), one
can no longer play with only one source because $S_2$ is 
the same for both copies of the quantum field, and,
therefore, both amplitudes in (1.54) respond. As a consequence,
all four matrices of second derivatives are generally involved:
\begin{equation}
\frac{\delta^2\ln Z}{\delta{\rm i}J_j
\delta{\rm i}J_k}\;,\quad
\frac{\delta^2\ln Z}{\delta{\rm i}J^*_j
\delta{\rm i}J^*_k}\;,\quad
\frac{\delta^2\ln Z}{\delta{\rm i}J^*_j
\delta{\rm i}J_k}\;,\quad
\frac{\delta^2\ln Z}{\delta{\rm i}J_j
\delta{\rm i}J^*_k}\;,
\end{equation}
i.e., the Green's function $G^{jk}$, its complex conjugate,
and two Wightman functions: 
$\langle1|{\hat\f}^j{\hat\f}^k|1\rangle$ and its transpose.
The Wightman functions can be expressed through $G^{jk}$
and the advanced or retarded Green's function:
\begin{equation}
{\rm i}\langle1|{\hat\f}^j
{\hat\f}^k|1\rangle
-{\rm i}\langle\f^j\rangle\langle\f^k\rangle=
G^{jk}-G^{+jk}+O(\hbar)=
G^{kj}-G^{-kj}+O(\hbar)\;.
\end{equation}

The result of the calculation is the following variational 
law for $G$:
\begin{equation}
\delta G=
G^-\delta S_2 G+
G\delta S_2 G^+-
G^-\delta S_2 G^+\;.
\end{equation}
It is no more the simple law (1.43) but it is, nevertheless,
universal because $G^+$ and $G^-$ are state-independent.
The variational law (1.65) is valid for any initial-value
problem.

The left-hand side of the expectation-value equations has
the form (1.45) as before but, since the variational law
for $G$ is different, the former inference about the
symmetry of $\delta\Gamma_i/\delta\f^j$ needs to be revised.
This inference is no longer valid. The advanced and retarded
Green's functions arrange it so that
\begin{equation}
\frac{\delta\Gamma_i(\f)}{\delta\f^j}=0\quad
\mbox{ when }i<j
\end{equation}
and
\begin{equation}
\frac{\delta\Gamma_i(\f)}{\delta\f^j}\ne0\quad
\mbox{ when }i>j\;.
\end{equation}
It follows that there is no action generating the 
expectation-value equations.

The nonexistence of an action for the initial-value problem
is seen also from the consideration of the Legendre transform
of the generating function (1.54). It is now a function of two
fields:
\begin{equation}
\Gamma(\f^*,\f)=\frac{1}{{\rm i}}\ln Z(J^*,J) - \f J 
+ \f^* J^*
\end{equation}
where
\begin{equation}
\f=\frac{\delta\ln Z(J^*,J)}{\delta{\rm i}J}\;,\qquad
\f^*=-\frac{\delta\ln Z(J^*,J)}{\delta{\rm i}J^*}\;.
\end{equation}
The expectation-value equations are obtained as
\begin{equation}
\f=\langle1|{\hat\f}|1\rangle\;:\qquad
\left.\frac{\delta\Gamma(\f^*,\f)}{\delta\f^i}
\right|_{\f^*=\f}=-J_i\;,
\hphantom{\left.\frac{\delta\Gamma(\f^*,\f)}{\delta\f^i}
\right|_{\f^*=\f}}
\end{equation}
and, therefore,
\begin{equation}
\Gamma_i(\f)=
\left.\frac{\delta\Gamma(\f^*,\f)}{\delta\f^i}
\right|_{\f^*=\f}\;.
\end{equation}
This is {\it not} a gradient.
}

%% file: lecture2.tex
\section[Lecture 2]{The In-Vacuum State and Schwinger--Keldysh diagrams}
\label{sec:2}
{\renewcommand{\theequation}{2.\arabic{equation}}
\subsubsection{Specification of The State}
In order to proceed, I need to specify the state. This will be
done in several steps.
\subparagraph{Step 1.} 
It will be assumed that $S_2$ is a second-order hyperbolic operator,
and the energy-momentum tensor of the field of small disturbances
$\delta\f^i$ with the action
\begin{equation}
\frac{1}{2}S_{ij}\delta\f^i\delta\f^j
\end{equation}
satisfies the dominant energy condition.
\subparagraph{Step 2.} 
The initial-value surface will be shifted to the remote past:
\begin{equation}
\Sigma_1\to -\infty\;.
\end{equation}
Consider the operator field equations (1.2)--(1.3):
\begin{equation}
J_i+S_i(c)+S_{ij}(c)({\hat\f}-c)^j+
\sum^{\infty}_{n=2}\frac{1}{n!}S_{ij_1\cdots j_n}(c)
({\hat\f}-c)^{j_1}\ldots ({\hat\f}-c)^{j_n}=0\;.
\end{equation}
If $c^i$ is some classical solution:
\begin{equation}
S_i(c)=-J_i\;,
\end{equation}
and ${\hat\phi}^i$ is an operator solution of $S_2$ against
the background $c^i$:
\begin{equation}
S_{ij}(c){\hat\phi}^j=0\;,
\end{equation}
then the field
\begin{equation}
{\hat\f}^i=c^i+{\hat\phi}^i\;,\qquad i\in\Sigma\to -\infty
\end{equation}
solves the operator dynamical equations asymptotically in the
remote past. It is a property of $S_2$ that its
solution with smooth data having a compact support or
decreasing at the spatial infinity 
decreases also in the timelike directions. Then, as 
$i\in\Sigma\to -\infty$,
the nonlinear terms in (2.3) decrease
even faster and are negligible. Thus, to build a Hilbert space
of states, it suffices to build a representation of the
algebra of ${\hat\phi}$'s.
\subparagraph{Step 3.} 
A Fock space will be built associated with the linear field
${\hat\phi}^i$. This amounts to expanding ${\hat\phi}^i$ in
some basis of solutions of $S_2(c)$:
\begin{equation}
S_2(c)\chi_A=0\;,
\end{equation}
\begin{equation}
{\hat\phi}^i=\chi_A^i{\hat a}_{\mbox{\scriptsize in}}{}^A+
\overline{\chi}_A^i{\hat a}^{+}_{\mbox{\scriptsize in}}{}^A
\end{equation}
where the overline means complex conjugation, and the basis functions
$\chi_A^i$ are normalized with the aid of the inner product:
\begin{equation}
(\chi_A,\chi_B)=0\;,\qquad
(\overline{\chi}_A,\chi_B)=\delta_{AB}\;,
\end{equation}
\begin{equation}
(\phi_1,\phi_2)\equiv -{\rm i}\int\limits_\Sigma
\phi_1W_\mu\phi_2\,d\Sigma^\mu\;.
\end{equation}
Here $W_\mu$ is the Wronskian of $S_2$. In this way, the concept
is introduced of {\it some} particles detectable in the past.
What kind of particles are these, i.e.,
what kind of detectors detect these particles -- depends on the
choice of the basis of solutions but, in any case, the
following functions will be chosen for the local observables $Q$:
\begin{equation}
Q^A({\hat\f}\Bigl|_\Sigma)=
-{\rm i}\delta^{AB}
\int\limits_\Sigma \chi_BW_\mu({\hat\f}-c)\,d\Sigma^\mu\;,
\end{equation}
$$
\Sigma\to -\infty\;.
$$
One needs these observables only on the initial-value surface, and,
there, they coincide with the annihilation operators of the introduced
particles:
\begin{equation}
Q^A({\hat\f}\Bigl|_{\Sigma\to -\infty})=
{\hat a}_{\mbox{\scriptsize in}}{}^A\;.
\end{equation}
The choice of the quantum state will be made in favour of
the zero-eigenvalue eigenstate of these observables:
\begin{equation}
{\hat a}_{\mbox{\scriptsize in}}{}^A|1\rangle=0\;.
\end{equation}
This is the vacuum of the introduced particles.

It follows from (2.6) and (2.8) that the
field's expectation value in the state (2.13), when taken in the remote
past, coincides with the classical solution $c^i$:
\begin{equation}
\langle1|{\hat\f}^i|1\rangle=c^i\;,\qquad i\in\Sigma\to -\infty\;.
\end{equation}
The ad hoc classical solution $c^i$ can then be eliminated completely
both from the asymptotic form of the quantum field
\begin{equation}
{\hat\f}^i=\langle\f^i\rangle+{\hat\phi}^i\;,\qquad i\in\Sigma\to -\infty
\end{equation}
and from the equation defining the Fock modes
\begin{equation}
S_{ij}(\langle\f\rangle){\hat\phi}^j=0\;,\qquad i\in\Sigma\to -\infty\;.
\end{equation}
Only the mean field itself figures as a background.

The specification of the state is, however, not completed
because the mean field in the past remains an arbitrary classical
solution:
\begin{equation}
S_i(\langle\f\rangle)=-J_i
\;,\qquad i\in\Sigma\to -\infty
\end{equation}
and the state itself remains the vacuum of undefined particles.
To make the final determination, one more step is needed.
\subparagraph{Step 4.} 
The final choice of the state assumes one more limitation
on the original action. Namely, it will be assumed that the
external source $J_i$ and all the external fields that may
be present in the action $S$ are asymptotically static in the past.
This means that, asymptotically in the past, there exists a
vector field $\xi^\mu$ such that it is nowhere tangent to any of the
Cauchy surfaces, and the Lie derivative in the direction
of $\xi^\mu$ of all external fields is zero. Specifically,
\begin{equation}
{\cal L}_\xi J_i=0\;,\qquad i\in\Sigma\to -\infty\;.
\end{equation}

If this limitation is fulfilled, then, among the 
solutions of (2.17) for the mean field in the past, there is the static
one:
\begin{equation}
{\cal L}_\xi \langle\f^i\rangle=0\;,
\qquad i\in\Sigma\to -\infty\;.
\end{equation}
Choose it. Next, use the fact that, with this choice, the operator
$S_2(\langle\f\rangle)$ 
commutes with the Lie derivative, and choose for the basis solutions of
$S_2(\langle\f\rangle)$ the functions that, asymptotically in
the past, are eigenfunctions of the Lie derivative:
\begin{equation}
{\rm i}{\cal L}_\xi \chi_A^i=\varepsilon_A \chi_A^i\;,
\quad \varepsilon_A>0\;,
\quad i\in\Sigma\to -\infty\;.
\end{equation}
This fixes both the initial conditions for the mean field and
the type of particles whose vacuum is the chosen state. These
are particles with definite energies.

Since $S_2$ is a second-order hyperbolic operator, it contains
some tensor field, $g^{\mu\nu}$, contracting the second
derivatives. The inverse matrix, $g_{\mu\nu}$, can serve and does serve
in every respect as a metric on the base manifold. The metric
enters the original action $S$ either as a part of the quantum 
field ${\hat\f}^i$ or as an external field. In both cases it
is subject to equation (2.19). When applied to the metric, this is
the Killing equation. Thus, we assume the existence, asymptotically
in the past, of a timelike Killing vector $\xi^\mu$.

The specification of the quantum initial data is now
completed. The notation for the state defined above is
\begin{equation}
|1\rangle=|\mbox{in vac}\rangle\;,
\end{equation}
and its full name is relative standard in-vacuum state. It is "relative"
because it is relative to the background generated by an asymptotically 
static source. It is "standard" because it refers to the
standard concept of particles. It is "in" because these particles
are incoming. And it is "vacuum" because these particles are
absent.

The state should not necessarily be chosen as the zero-eigenvalue
eigenstate. Since the expectation-value equations do not depend
on the eigenvalues, they will have the same form for any
eigenstate of the annihilation operators, i.e., for any coherent
state
\begin{equation}
{\hat a}_{\mbox{\scriptsize in}}{}^A
|\mbox{in }\alpha\rangle=\alpha^A
|\mbox{in }\alpha\rangle\;.
\end{equation}
Only the initial conditions for the mean field will be different:
\begin{equation}
\langle\alpha\mbox{ in}|{\hat\f}^i
|\mbox{in }\alpha\rangle=c^i+\chi_A^i\alpha^A+
\overline{\chi}_A^i\overline{\alpha}^A\;,
\qquad i\in\Sigma\to -\infty\;.
\end{equation}
In addition to the static background $c^i$ generated by a source,
the mean field in the past contains now the incoming wave of an
arbitrary profile. This is the general setting of the classical
evolution problem for an observable field like the electromagnetic or
gravitational field. The fact that the nature of the state
has changed from classical to quantum did not affect this setting.

It will be useful to keep comparing the initial-value problem
with the boundary-value problem. In the latter case, one can
define similarly the out-vacuum state and specify the quantum
boundary data as
\begin{equation}
|1\rangle=|\mbox{in vac}\rangle\;,\qquad
|2\rangle=|\mbox{out vac}\rangle\;.
\end{equation}
\subsubsection{Perturbation Theory}
With this specification of the states, let us come back to the
mean-field equations. There remains to be obtained the Green's
function $G(\f)$ that figures in the loops. We need it for
an arbitrary background $\f$ but we have a variational law,
(1.43) or (1.65), which may be regarded as a differential equation
for $G(\f)$ with respect to $\f$. The only thing that is
missing and that depends on the choice of states is the
initial condition to this equation. It suffices, therefore, 
to know $G$ for only one background.

Then let us do the simplest: perturbation theory around the
trivial background. A second-order hyperbolic operator with
the trivial background is the D'Alembert operator with
flat metric, $\Box_0$:
\begin{equation}
S_2(\f)=\Box_0+P\;.
\end{equation}
The remainder is a perturbation $P$.

In the case of the boundary-value problem, the variational law is
(1.43), and, therefore, the expansion of $G(\f)$ is of the form
\begin{equation}
G(\f)=G_0+G_0PG_0+G_0PG_0PG_0+\ldots
\end{equation}
where $G_0$ is $G$ for the trivial background. This expansion is
to be inserted in the loop in the mean-field
equations
\begin{equation}
\frac{1}{2{\rm i}}S_{ijk}(\varphi)G^{jk}(\varphi)={}
\parbox{30pt}{
\begin{picture}(30,20)
\thicklines
\put(20,10){\circle{20}}
\put(0,10){\line(1,0){10}}
\put(0,12){$\scriptstyle i$}
\end{picture}
}\;\,.
\end{equation}
Let for simplicity $P$ be a potential. One obtains the loop
expanded in powers of $P$:
\begin{equation}
\parbox{30pt}{
\begin{picture}(30,20)
\thicklines
\put(20,10){\circle{20}}
\put(0,10){\line(1,0){10}}
\put(0,13){$x$}
\end{picture}
}
{}=\int dy_1\ldots dy_n\,F(x|y_1,\ldots y_n)P(y_1)\ldots P(y_n)\;. 
\end{equation}
The coefficients $F$ will be called formfactors.
The formfactors are loop diagrams
\begin{equation}
F(x|y)={}
\parbox{57pt}{
\begin{picture}(57,30)(-8,5)
\qbezier(0,20)(20,40)(40,20)
\qbezier(0,20)(20,0)(40,20)
\put(41,23){$y$}
\put(0,23){\llap{$x$}}
\end{picture}
}\;,
\end{equation}
\begin{equation}
F(x|y_1,y_2)={}
\parbox{53pt}{
\begin{picture}(53,54)(-8,-7)
\put(0,20){\line(3,2){30}}
\put(0,20){\line(3,-2){30}}
\put(30,0){\line(0,1){40}}
\put(0,23){\llap{$x$}}
\put(33,39){$y_1$}
\put(33,-2){$y_2$}
\end{picture}
}\;,
\end{equation}
$$
\hbox to 120pt{{}\dotfill{}}
$$
with the same propagator for all lines: the trivial-background
Green's function
\begin{equation}
\parbox{30pt}{
\begin{picture}(30,3)
\put(0,1.5){\line(1,0){30}}
\end{picture}
}
{}=G_0\;.
\end{equation}
What is $G_0$? With the trivial background and the standard
in- and out- vacuum states, it is the Feynman Green's function:
\begin{equation}
G_0=G_{\mbox{\sc\scriptsize feynman}}\;.
\end{equation}

Let us do the same thing for the initial-value problem. The loop
in the expectation-value equations will, in the same way, be
expanded in powers of the perturbation, and the expansion will
have the same form (2.28), but the formfactors will be different
because the variational law for $G$ is different. It is now
(1.65) rather than (1.43). Using this law, one obtains for the
formfactors three diagrams in place of one:
\begin{equation}
F(x|y)={}
\parbox{57pt}{
\begin{picture}(57,30)(-8,5)
\qbezier(0,20)(20,40)(40,20)
\qbezier(0,20)(20,0)(40,20)
\put(21.5,30){\vector(-1,0){3}}
\put(41,23){$y$}
\put(0,23){\llap{$x$}}
\end{picture}
}
{}+{}
\parbox{57pt}{
\begin{picture}(57,30)(-8,5)
\qbezier(0,20)(20,40)(40,20)
\qbezier(0,20)(20,0)(40,20)
\put(21.5,10){\vector(-1,0){3}}
\put(41,23){$y$}
\put(0,23){\llap{$x$}}
\end{picture}
}
{}-{}
\parbox{57pt}{
\begin{picture}(57,30)(-8,5)
\qbezier(0,20)(20,40)(40,20)
\qbezier(0,20)(20,0)(40,20)
\put(21.5,30){\vector(-1,0){3}}
\put(21.5,10){\vector(-1,0){3}}
\put(41,23){$y$}
\put(0,23){\llap{$x$}}
\end{picture}
}\;,
\end{equation}
five diagrams in place of one:
\begin{eqnarray}
F(x|y_1,y_2)={}
\parbox{53pt}{
\begin{picture}(53,54)(-8,-7)
\put(0,20){\line(3,2){30}}
\put(30,0){\vector(-3,2){15}}
\put(0,20){\line(3,-2){15}}
\put(30,40){\vector(0,-1){20}}
\put(30,0){\line(0,1){20}}
\put(0,23){\llap{$x$}}
\put(33,39){$y_1$}
\put(33,-2){$y_2$}
\end{picture}
}
{}+{}
\parbox{53pt}{
\begin{picture}(53,54)(-8,-7)
\put(30,40){\vector(-3,-2){15}}
\put(0,20){\line(3,2){15}}
\put(30,0){\vector(-3,2){15}}
\put(0,20){\line(3,-2){15}}
\put(30,0){\line(0,1){40}}
\put(0,23){\llap{$x$}}
\put(33,39){$y_1$}
\put(33,-2){$y_2$}
\end{picture}
}
{}+{}
\parbox{53pt}{
\begin{picture}(53,54)(-8,-7)
\put(30,40){\vector(-3,-2){15}}
\put(0,20){\line(3,2){15}}
\put(0,20){\line(3,-2){30}}
\put(30,40){\line(0,-1){20}}
\put(30,0){\vector(0,1){20}}
\put(0,23){\llap{$x$}}
\put(33,39){$y_1$}
\put(33,-2){$y_2$}
\end{picture}
}
\nonumber\\
{}-{}
\parbox{53pt}{
\begin{picture}(53,54)(-8,-7)
\put(30,40){\vector(-3,-2){15}}
\put(0,20){\line(3,2){15}}
\put(30,0){\vector(-3,2){15}}
\put(0,20){\line(3,-2){15}}
\put(30,40){\vector(0,-1){20}}
\put(30,0){\line(0,1){20}}
\put(0,23){\llap{$x$}}
\put(33,39){$y_1$}
\put(33,-2){$y_2$}
\end{picture}
}
{}-{}
\parbox{53pt}{
\begin{picture}(53,54)(-8,-7)
\put(30,40){\vector(-3,-2){15}}
\put(0,20){\line(3,2){15}}
\put(30,0){\vector(-3,2){15}}
\put(0,20){\line(3,-2){15}}
\put(30,40){\line(0,-1){20}}
\put(30,0){\vector(0,1){20}}
\put(0,23){\llap{$x$}}
\put(33,39){$y_1$}
\put(33,-2){$y_2$}
\end{picture}
}\;,
\end{eqnarray}
and so on. There are two types of propagators in these diagrams:
the trivial-background $G$, and the trivial-background retarded
or advanced Green's function. Respectively, there are two types
of lines:
\begin{equation}
\parbox{30pt}{
\begin{picture}(30,3)
\put(0,1.5){\line(1,0){30}}
\end{picture}
}
{}=G_0\;,\qquad
\parbox{30pt}{
\begin{picture}(30,3)
\put(30,1.5){\vector(-1,0){30}}
\end{picture}
}
{}=G_0^-\mbox{ or }G_0^+\;.
\end{equation}
In the latter case, the arrow
points the direction of growth of time. And what is now $G_0$?
In terms of the linear field (2.5) it is
\begin{equation}
\frac{1}{{\rm i}}G_0^{jk}=\langle\mbox{in vac}|
\overleftarrow{T}({\hat\phi}^j{\hat\phi}^k)
|\mbox{in vac}\rangle
\Bigl|_{\mbox{\scriptsize trivial background}}
\end{equation}
and differs from the previous case in that the
"$\langle\mbox{out vac}|$" is replaced by the
"$\langle\mbox{in vac}|$". But, with the trivial background, the
vacuum for the linear field is stable. The out-vacuum coincides
with the in-vacuum. Therefore,
\begin{equation}
G_0=G_{\mbox{\sc\scriptsize feynman}}\quad\mbox{(again!)}\;.
\end{equation}

The diagrams above are called Schwinger--Keldysh diagrams.
There is not more than one Feynman propagator in every diagram.
The remaining ones are the retarded and advanced Green's
functions organized in a special way and with special signs of
the diagrams themselves. There is a mystery in this special
arrangement. What do these diagrams want
to tell us? We must disclose their secret because working with
them directly is not what can be recommended.
\subsubsection{Mystery of The Schwinger--Keldysh Diagrams}
One thing is obvious right away. In the diagrams above, there
is always a chain of retarded Green's functions connecting
a given point $y$ with the observation point $x$. Therefore,
the formfactor vanishes if at least one of the $y$'s is in
the future of $x$. This is the {\it retardation property}
\begin{equation}
F(x|y_1,\ldots y_n)=0\quad\mbox{ when }\; y_m>x\;,\;\;\forall m\;.
\end{equation}
But this is true of every Schwinger--Keldysh diagram, and why do they
appear in the special combinations? What is the role of the
Feynman propagator?

Let us make a Fourier transformation of the formfactor with
respect to the differences $(x-y_m)$ in the Minkowski coordinates:
\begin{equation}
F(x|y_1,\ldots y_n) 
=\int dk_1\ldots dk_n\,\exp\Bigl({\rm i}\sum_{m=1}^n k_m(x-y_m)\Bigr)
f(k_1,\ldots k_n)\;. 
\end{equation}
How come that $F$ possesses the retardation property? It is
only that $f$ should admit an analytic continuation
to the upper half-plane in the timelike components of $k$'s.
Then, for $y_m$ later than $x$, we shall be able to close the
integration contour in the upper half-plane of $k_m^0$,
and the integral will vanish. There should be a function
of complex momenta
$f(z_1,\ldots z_n)$ analytic in the upper
half-planes of $z_m^0$ and such that 
$f(k_1,\ldots k_n)$ 
is its limiting value on the real axes:
\begin{equation}
f(k_1,\ldots k_n)=f(z_1,\ldots z_n)\Bigl|_{\textstyle z_m^0=k_m^0+
{\rm i}\varepsilon}\;. 
\end{equation}
Let us build this function.

All diagrams in a given-order formfactor are similar. They all
are integrals over the momentum circulating in the loop, and
the integrands are identical. The difference is only in the
integration contours. Thus any diagram in the lowest-order
formfactor $f(k)$ is of the form
\begin{equation}
\parbox{57pt}{
\begin{picture}(57,30)(-8,5)
\qbezier(0,20)(20,40)(40,20)
\qbezier(0,20)(20,0)(40,20)
\put(0,20){\circle*{1.5}}
\put(41,23){$k$}
\end{picture}
}
{}=\int d{\vec p}\int\limits_{\displaystyle {\cal C}}dp^0\,
\frac{\mbox{polynomial in momenta}}{\left(-p^0{}^2+{\vec p}^2\right)
\left(-(p^0-k^0)^2+({\vec p}-{\vec k})^2\right)}\;.
\end{equation}
There are, generally, as many factors in the denominator as
there are propagators in the loop, and each factor contains
two poles. The contour ${\cal C}$ passes round them in
accordance with the type of the propagator. One of the three
rules applies to each pair of poles:
$$
\begin{array}{c}
\parbox{164pt}{
\begin{picture}(164,18)(-28,-12)
\put(-28,0){\line(1,0){10}}
\put(18,0){\line(1,0){10}}
\put(-12,0){\circle*{3}}
\put(12,0){\circle*{3}}
\put(0,0){\oval(36,12)[t]}
\put(50,0){retardation rule,}
\end{picture}
}\\
\parbox{164pt}{
\begin{picture}(164,24)(-28,-12)
\put(-28,0){\line(1,0){10}}
\put(18,0){\line(1,0){10}}
\put(-12,0){\circle*{3}}
\put(12,0){\circle*{3}}
\put(0,0){\oval(36,12)[b]}
\put(50,0){advancement rule,}
\end{picture}
}\\
\parbox{164pt}{
\begin{picture}(164,18)(-28,-6)
\put(-28,0){\line(1,0){10}}
\put(18,0){\line(1,0){10}}
\put(-12,0){\circle*{3}}
\put(12,0){\circle*{3}}
\put(-12,0){\oval(12,12)[b]}
\put(12,0){\oval(12,12)[t]}
\put(-6,0){\line(1,0){12}}
\put(50,0){Feynman rule.}
\end{picture}
}\\
\end{array}
$$

Let us now shift the external momentum $k^0$ to the complex plane.
The poles will shift to the complex plane but we shall also
deform smoothly the contour so that it do not cross the poles.
In this way one can build a function of complex momenta for
each Schwinger--Keldysh diagram. 
Thus the lowest-order formfactor with
complex momentum, $f(z)$, is a sum of three functions:
\begin{equation}
f(z)=
\int d{\vec p}\int\limits_{\displaystyle {\cal C}_1}dp^0\,(\ldots)+
\int d{\vec p}\int\limits_{\displaystyle {\cal C}_2}dp^0\,(\ldots)-
\int d{\vec p}\int\limits_{\displaystyle {\cal C}_3}dp^0\,(\ldots)\;,
\end{equation}
and the contours ${\cal C}_1$, ${\cal C}_2$, ${\cal C}_3$
for $z^0$ in the upper half-plane are shown in Fig. 1.
By considering the pinch
conditions, i.e., the conditions that the poles pinch the
integration contour, one can check in each case that these
functions can have singularities only on the real axis.
Therefore, if we consider them in the upper half-plane, they
are analytic, and their limits on the real axis are our
original diagrams. 

There remains to be understood what are these functions.
Since the integrands are identical, the sum of the integrals in (2.42)
is the integral over the sum of the contours
\begin{equation}
f(z)=
\int d{\vec p}\int\limits_{\displaystyle {\cal C}_1+{\cal C}_2-{\cal C}_3}
dp^0\,(\ldots)\;.
\end{equation}
Sum up the three contours in Fig. 1. The resultant contour is
such that every pair of poles is passed round by the Feynman rule.
It may be called Feynman contour.
\input figone.tex

But the Feynman contour defines also the in-out formfactor (2.29)
in which both propagators are Feynman, except that the in-out
formfactor is not the limit of $f(z)$ from the upper half-plane.
It is this limit on only half of the real axis, and on the
other half it is the limit from the lower half-plane.
The \mbox{in-in} and in-out formfactors are different boundary values of
the same complex function having a cut on the real axis:
\begin{equation}
\mbox{in-in}\;:\qquad f(k)=f(z)
\Bigl|_{\textstyle z^0=k^0+{\rm i}\varepsilon}\;,
\hphantom{\;{}}
\end{equation}
\begin{equation}
\mbox{in-out}\;:\qquad f(k)=f(z)
\Bigl|_{\textstyle z^0=(1+{\rm i}\varepsilon)k^0}\;,
\end{equation}
and the function itself is the integral over the Feynman contour
\begin{equation}
f(z)=
\int d{\vec p}\int\limits_{\displaystyle
{\cal C}_{\mbox{\sc\scriptsize feynman}}}
dp^0\,(\ldots)\;.
\end{equation}

The same is true of all $n$-th order formfactors, and this
is a disclosure of the mystery. In each case, the set of
Schwinger--Keldysh diagrams is just a splitting of one
Feynman diagram whose purpose is to display the retardation
property and in this way to tell us which boundary value
is to be taken.
\subsubsection{Reduction to The Euclidean Effective Action}
The Feynman contour is famous for the fact that, when the
external momenta are on the imaginary axis, the Feynman contour
is the imaginary axis itself. With all the momenta imaginary,
both the external ones and the one circulating in the loop, this is the
Euclidean formfactor. Then we can {\it start} with the  
calculation of the Euclidean
formfactor and next analytically continue it in momenta from the
imaginary axis to the real axis either in the way shown in Fig. 2(a)
or in the way shown in Fig. 2(b). In the first case we shall obtain
the in-out formfactor, and in the second case the in-in formfactor
of Lorentzian theory. It is invaluable that loops can be calculated
Euclidean.
\input figtwo.tex

Then let us make one more step. A formfactor with the Euclidean momentum
can be put in the spectral form
\begin{equation}
f(k)=\int\limits_0^\infty dm^2\,
\frac{\rho(m^2)}{m^2+k^2}+\mbox{ a polynomial in }k^2\;,
\end{equation}
$$
k^2>0
$$
with some spectral weight $\rho(m^2)$, the resolvent $1/(m^2+k^2)$,
and a polynomial accounting for a possible growth of $f(k)$ at
$k^2\to\infty$. There are similar forms for the higher-order formfactors. 
If the formfactor is in the spectral form, the procedure of analytic
continuation boils down merely to replacing the Euclidean resolvent
with the retarded or Feynman resolvent:
\begin{equation}
\hphantom{{}\;{}}
\mbox{in-in}\;:\qquad 
f(k)=\int\limits_0^\infty dm^2\,
\frac{\rho(m^2)}{m^2-(k^0+{\rm i}\varepsilon)^2+{\vec k}^2}
+\mbox{ a polynomial in }k^2\;,
\end{equation}
\begin{equation}
\mbox{in-out}\;:\qquad
f(k)=\int\limits_0^\infty dm^2\,
\frac{\rho(m^2)}{m^2-k^0{}^2+{\vec k}^2-{\rm i}\varepsilon}+
\mbox{ a polynomial in }k^2\;.
\end{equation}
Note that the spectral weight is the same in all cases: the one
of the Euclidean loop. Thus, the problem boils down to obtaining
the spectral weights of the Euclidean formfactors.

Then back from the Fourier-transformed formfactors to the
formfactors themselves, and from the formfactors to the
mean-field equations. For the loop in these equations
expanded in powers of the perturbation, we obtain an
expression of the following form:
\begin{eqnarray}
\parbox{30pt}{
\begin{picture}(30,20)
\thicklines
\put(20,10){\circle{20}}
\put(0,10){\line(1,0){10}}
\put(0,13){$x$}
\end{picture}
}
&{}=&(c_1+c_2\Box_0+\ldots)P(x)\nonumber\\
&{}+&\int\limits_0^\infty dm^2\,\rho(m^2)
\frac{1}{m^2-\Box_0}P(x)\nonumber\\
&{}+&\int\limits_0^\infty dm_1^2dm_2^2dm_3^2\,\rho(m_1^2,m_2^2,m_3^2)
\nonumber\\
&&{}\times\frac{1}{m_1^2-\Box_0}\left[
\left(\frac{1}{m_2^2-\Box_0}P(x)\right)
\left(\frac{1}{m_3^2-\Box_0}P(x)\right)\right]\nonumber\\
&{}+&\ldots\;.
\end{eqnarray}
Here the first term is local. It comes from the polynomial in
the spectral form. The remaining terms are nonlocal but expressed
through the resolvent which is a Green's function of the massive
operator $\Box_0-m^2$. It is initially the Euclidean Green's function
since we are calculating the Euclidean loop. For the Lorentzian equations,
we arrive at the following rule.
To obtain the expectation-value equations in the in-vacuum state,
replace all the Euclidean resolvents in (2.50) with the retarded
Green's functions. 
To obtain the mean-field equations for the
in-out problem, replace all the Euclidean resolvents
with the Feynman Green's functions:
\begin{equation}
\parbox{172.7pt}{
\begin{picture}(172.7,57.6)(-67.1,-24.3)
\put(0,0){\vector(1,0){48.6}}
\put(0,0){\vector(2,1){48.6}}
\put(0,0){\vector(2,-1){48.6}}
\put(55.6,24.3){Euclidean,}
\put(55.6,0){Retarded,}
\put(55.6,-24.3){Feynman.}
\put(-11.4,0){\llap{All ${\displaystyle\frac{1}{m^2-\Box_0}}$}}
\end{picture}
}
\end{equation}

At every level of expectation-value theory, there are proofs
that the expectation-value equations possess two basic properties:
they are real and causal. Causality is the retardation property
discussed above. But it is not enough to have proofs. These
properties should be manifestly built into the working formalism.
Expression (2.50) offers such a formalism. Since the retarded
resolvent secures the causality and is real, this expression
is manifestly real and causal.

But even this is not enough. The theory may possess symmetries,
and one may want these symmetries to be manifest. To this end
it will be noted that, although expansion (2.50) is obtained in
terms of the trivial-background resolvent $1/(m^2-\Box_0)$,
it can be regrouped so as to restore the full-background
resolvent
\begin{equation}
\frac{1}{m^2-S_2}=\frac{1}{m^2-\Box_0-P}
\end{equation}
at each order. It does not matter whether this regrouping will
be made in the expectation-value equations or in the Euclidean
equations because the retarded and Euclidean Green's functions
obey the same variational law (1.43):
\begin{equation}
\frac{1}{m^2-\Box_0}=\frac{1}{m^2-S_2}-
\frac{1}{m^2-S_2}P\frac{1}{m^2-S_2}+\ldots\;.
\end{equation}
This proves that the rule of replacing resolvents applies to
the full-background resolvents as well as to the trivial-background
ones. The latter fact is important because the Euclidean loops
can be calculated covariantly from the outset, and the transition
to the expectation-value equations by replacing
the full-background resolvents does not break 
the manifest symmetries. The expectation-value equations are
obtained in as good an approximation as the Euclidean equations
are.

There remains to be made a final observation. For the Euclidean
equations, {\it there is} an effective action: 
\begin{equation}
\parbox{30pt}{
\begin{picture}(30,20)
\thicklines
\put(20,10){\circle{20}}
\put(0,10){\line(1,0){10}}
\put(0,12){$\scriptstyle i$}
\end{picture}
}
{}=
\frac{\delta}{\delta\varphi^i}
\;
\parbox{20pt}{
\begin{picture}(20,20)
\thicklines
\put(10,10){\circle{20}}
\end{picture}
}
\end{equation}
because the variational law for the Euclidean Green's function is (1.43).
It is invaluable that loops can be calculated without external
lines. This reduces the calculations greatly, helps to control
symmetries, helps to control renormalizations.

Thus, at the end of the day, we conclude that {\it there is} an
action that generates the expectation-value equations but it
does so indirectly, i.e., {\it not} through the least-action
principle. To make this clear, consider (for the illustrative
purposes only) any quadratic action:
$$
\Gamma(\f)=\frac{1}{2}\int dx\,\f f(\Box_0)\f\;.
$$
Whatever the operator $f(\Box_0)$ is, in the variational derivative
it gets symmetrized:
$$
\frac{\delta\Gamma(\f)}{\delta\f}=\frac{1}{2}\left(f(\Box_0)+
f^{\rm T}(\Box_0)\right)\f=f^{\mbox{\small sym}}
(\Box_0)\f\;.
$$
Assuming that the function $f(\Box_0)$ is in the spectral form
$$
f(\Box_0)=\int\limits_0^\infty dm^2\,\rho(m^2)
\frac{1}{m^2-\Box_0}\;,
$$
one obtains the variational equations with the symmetrized resolvent:
$$
\int\limits_0^\infty dm^2\,\rho(m^2)
\left(\frac{1}{m^2-\Box_0}\right)^{\mbox{sym}}\f=-J\;.
$$
These cannot be the expectation-value equations since they are
not causal. But, through the derivation above, we know how to
correct this: just to replace the symmetrized resolvent with the
retarded resolvent. The corrected equations
$$
\int\limits_0^\infty dm^2\,\rho(m^2)
\left(\frac{1}{m^2-\Box_0}\right)^{\mbox{ret}}\f=-J\;.
$$
do not already follow from any action although indirectly they do.
Only if the action $\Gamma(\f)$ is local, i.e., 
the function $f(\Box_0)$ is polynomial,
the least-action principle holds directly.

Two precepts should be kept in mind when using the formalism above.
First, the replacement rule concerns the resolvents of the formfactors
and not the
propagators in the loop. The loop should be calculated Euclidean.
Hence
\subparagraph{First Precept:} 
first do the loop, next replace the resolvents.\par
\medskip
\noindent Second, the replacement of resolvents
is to be made in the equations and not in the action. It does not
make sense to make it in the action. Hence
\subparagraph{Second Precept:} 
first vary the action, next replace the resolvents.\par
\medskip
We thus go over to the calculation of the Euclidean effective action.
}

%% file: figone.tex
\begin{figure}
\centering
\begin{picture}(236,248)(-78,-200)
\put(0,-130){\line(0,1){169}}
\put(0,-200){\line(0,1){35}}
\put(-12,0){\circle*{3}}
\put(-12,-60){\circle*{3}}
\put(-12,-120){\circle*{3}}
\put(-12,-190){\circle*{3}}
\put(12,0){\circle*{3}}
\put(12,-60){\circle*{3}}
\put(12,-120){\circle*{3}}
\put(12,-190){\circle*{3}}
\put(70,19){\circle*{3}}
\put(70,-41){\circle*{3}}
\put(70,-101){\circle*{3}}
\put(70,-171){\circle*{3}}
\put(94,19){\circle*{3}}
\put(94,-41){\circle*{3}}
\put(94,-101){\circle*{3}}
\put(94,-171){\circle*{3}}
{\thicklines
\put(-48,0){\line(1,0){30}}
\put(-48,-60){\line(1,0){30}}
\put(-48,-120){\line(1,0){30}}
\put(-48,-190){\line(1,0){30}}
\put(0,0){\oval(36,12)[t]}
\put(18,0){\line(1,0){70}}
\put(88,0){\line(0,1){19}}
\put(100,0){\line(0,1){19}}
\put(94,19){\oval(12,12)[t]}
\put(100,0){\vector(1,0){58}}
\put(-12,-60){\oval(12,12)[b]}
\put(12,-60){\oval(12,12)[t]}
\put(-6,-60){\line(1,0){12}}
\put(18,-60){\vector(1,0){140}}
\put(0,-120){\oval(36,12)[t]}
\put(18,-120){\vector(1,0){140}}
\put(-12,-190){\oval(12,12)[b]}
\put(12,-190){\oval(12,12)[t]}
\put(-6,-190){\line(1,0){12}}
\put(18,-190){\line(1,0){70}}
\put(88,-190){\line(0,1){19}}
\put(100,-190){\line(0,1){19}}
\put(94,-171){\oval(12,12)[t]}
\put(100,-190){\vector(1,0){58}}
\put(-78,0){\line(1,0){12}}
\put(-78,-60){\line(1,0){12}}
\put(-78,-120){\line(1,0){12}}
\put(-72,-6){\line(0,1){12}}
\put(-72,-66){\line(0,1){12}}
}
\put(158,6){\llap{${\cal C}_1$}}
\put(158,-54){\llap{${\cal C}_2$}}
\put(158,-114){\llap{${\cal C}_3$}}
\put(158,-184){\llap{${\cal C}_{\mbox{\sc\scriptsize feynman}}$}}
\put(-12,-155){{\bf SUM:}}
\put(3,39){\fbox{$p^0$ plane}}
\end{picture}
\caption{Integration contours for the three diagrams in the
lowest-order formfactor~(2.42). The sum of the contours is the Feynman
contour.}
\label{fig:1}
\end{figure}
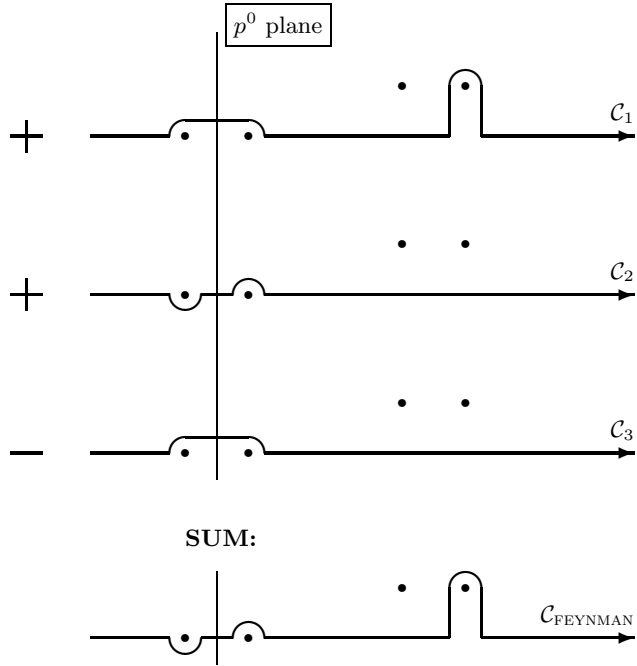

%% file: figtwo.tex
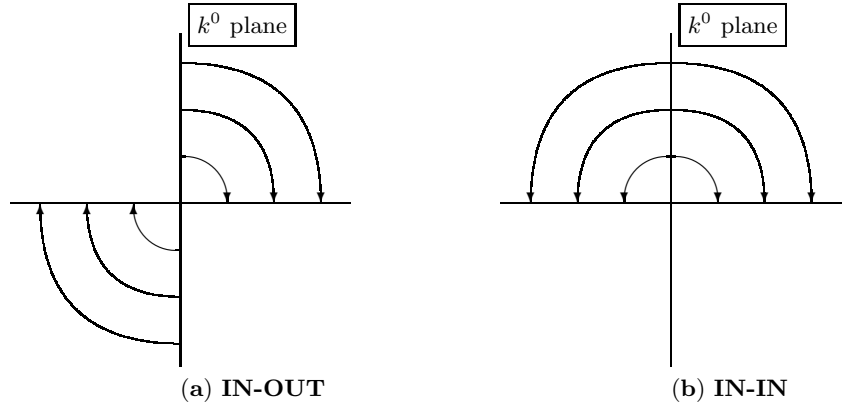
\begin{figure}
\centering
\begin{picture}(314.2,146.2)(-157.1,-72.7)
\put(28.5,0){\line(1,0){128.6}}
\put(-28.5,0){\line(-1,0){128.6}}
\put(92.8,-61.7){\line(0,1){126.2}}
\put(-92.8,-61.7){\line(0,1){126.2}}
\put(92.8,0){\oval(35.4,35.4)[t]}
\qbezier(92.8,35.4)(57.4,35.4)(57.4,0)
\qbezier(92.8,35.4)(128.2,35.4)(128.2,0)
\qbezier(92.8,53.1)(39.7,53.1)(39.7,0)
\qbezier(92.8,53.1)(145.9,53.1)(145.9,0)
\put(-92.8,0){\oval(35.4,35.4)[tr]}
\qbezier(-92.8,35.4)(-57.4,35.4)(-57.4,0)
\qbezier(-92.8,-35.4)(-128.2,-35.4)(-128.2,0)
\qbezier(-92.8,53.1)(-39.7,53.1)(-39.7,0)
\qbezier(-92.8,-53.1)(-145.9,-53.1)(-145.9,0)
\put(-92.8,0){\oval(35.4,35.4)[bl]}
\put(110.5,3){\vector(0,-1){3}}
\put(128.2,3){\vector(0,-1){3}}
\put(145.9,3){\vector(0,-1){3}}
\put(75.1,3){\vector(0,-1){3}}
\put(57.4,3){\vector(0,-1){3}}
\put(39.7,3){\vector(0,-1){3}}
\put(-75.1,3){\vector(0,-1){3}}
\put(-57.4,3){\vector(0,-1){3}}
\put(-39.7,3){\vector(0,-1){3}}
\put(-110.5,-3){\vector(0,1){3}}
\put(-128.2,-3){\vector(0,1){3}}
\put(-145.9,-3){\vector(0,1){3}}
\put(95.8,64.5){\fbox{$k^0$ plane}}
\put(-89.8,64.5){\fbox{$k^0$ plane}}
\put(92.8,-72.7){({\bf b}) {\bf IN-IN}}
\put(-92.8,-72.7){({\bf a}) {\bf IN-OUT}}
\end{picture}
\caption{Analytic continuation of the Euclidean formfactor 
that gives ({\bf a}) the in-out formfactor,
({\bf b}) the in-in formfactor of Lorentzian theory.}
\label{fig:2}
\end{figure}

%% file: lecture3.tex
\section[Lecture 3]{The Effective Action}
\label{sec:3}
{\renewcommand{\theequation}{3.\arabic{equation}}
\subsubsection{The Operator $S_2$}
The $\f^i$ is a set of fields for which a more explicit notation will
now be used:
\begin{equation}
\f^i=\f^a(x)\;.
\end{equation}
The operator $S_2$ acts on a small disturbance of $\f^i$ and is
a second-order differential operator
\begin{equation}
S_{ij}\delta\f^j=\left(X^{\mu\nu}_{ab}\partial_\mu\partial_\nu
+Y^\mu_{ab}\partial_\mu+Z_{ab}\right)\delta\f^b(x)\;.
\end{equation}
The generality of this operator will, however, be restricted
by the condition that the coefficient of the senior term factorizes as
\begin{equation}
X^{\mu\nu}_{ab}=\omega_{ab}\,g^{\mu\nu}\;,\qquad
\det\omega_{ab}\ne 0\;,\;\;\det g^{\mu\nu}\ne 0\;.
\end{equation}
In this case, the operator (3.2) is said to be diagonal, or minimal,
or nonexotic. Condition (3.3) is too restrictive and not necessary.
It can be replaced by a more general condition
\begin{equation}
\det\left(X^{\mu\nu}_{ab}n_\mu n_\nu\right)
=C(g^{\mu\nu}n_\mu n_\nu)^d
\quad\forall n_\mu\;,\;\quad d=\dim a\;,\;\;C\ne 0\;,\;\;
\det g^{\mu\nu}\ne 0\;,
\end{equation}
and even this condition can be generalized. Higher-order and
first-order operators can also be considered but, in all of these
cases, the Green's functions of $S_2$ are expressed through the
Green's functions of a diagonal second-order operator. The case
(3.3) is basic.

In the case (3.3), the matrix $\omega_{ab}$ can be factored out:
\begin{equation}
S_{ij}\delta\f^j=\omega_{ac} H^c_b\delta\f^b(x)\;,
\end{equation}
and a covariant derivative can be introduced:
\begin{equation}
\mathop{\nabla_\mu}\delta\f^a=\left(\delta^a_b\partial_\mu
+{\cal A}_\mu{}^a_b\right)\delta\f^b
\end{equation}
so as to absorb the first-order term:
\begin{equation}
H^a_b=\delta^a_b g^{\mu\nu}\nabla_\mu\nabla_\nu+P^a_b\;.
\end{equation}
This is the final form of $S_2$. A short notation will
be used:
\begin{equation}
H=\Box{\hat 1}+{\hat P}
\end{equation}
where
\begin{equation}
\Box\equiv g^{\mu\nu}\nabla_\mu\nabla_\nu\;,
\end{equation}
and the hat designates a matrix in $a,b$:
\begin{equation}
{\hat 1}=\delta^a_b\;,\quad
{\hat P}=P^a_b\;,\quad\mathop{\rm tr}{\hat P}=P^a_a\;,\quad 
\mbox{etc.}
\end{equation}

The matrix $\omega_{ab}$ may be regarded as a local metric in the space
of fields. The symmetry of $S_2$ implies that this matrix
is symmetric, covariantly constant, and converts ${\hat P}$ into a 
symmetric form:
\begin{equation}
\omega_{ab}=\omega_{ba}\;,\quad\;\;\nabla_\mu\omega_{ab}=0\;,
\end{equation}
\begin{equation}
P^c_a\omega_{cb}-P^c_b\omega_{ca}=0\;.
\end{equation}
The dominant energy condition implies that $\omega_{ab}$ is positive
definite. The matrix $g^{\mu\nu}$ is the inverse of the metric on
the base manifiold. Since we are considering Euclidean theory,
this metric is positive definite too.

Apart from the algebraic factor $\omega_{ac}$ in (3.5), the operator $S_2$
contains three background fields:
\begin{equation}
g^{\mu\nu}\;,\quad\nabla_\mu\;,\quad{\hat P}
\end{equation}
i.e., the metric, the connection (or covariant derivative), and the
matrix potential. And where is the original background $\f$ of
$S_2(\f)$? When $S_2$ is calculated from the action $S$, the metric,
connection, and potential are obtained as functions of the
original set of fields $\f$, but from now on it does not matter.
The effective action is expressed in a universal manner through
the fields (3.13) only.

The strengths of the fields (3.13) are respectively the Riemann
tensor, the commutator of covariant derivatives, and the potential
which is its own strength:
\begin{equation}
R_{\alpha\beta\mu\nu}\;,\quad
[\nabla_\mu,\nabla_\nu]={\hat{\cal R}}_{\mu\nu}\;,\quad {\hat P}\;.
\end{equation}
I shall call these field strengths curvatures and use for them
the collective notation
\begin{equation}
\left(\,R_{\alpha\beta\mu\nu}\,,\;
{\hat{\cal R}}_{\mu\nu}\,,\; {\hat P}\,\right)=\Re\;.
\end{equation}
The following contractions of the curvatures will be called
currents:
\begin{equation}
{\hat J}_\mu\equiv\nabla^\nu{\hat{\cal R}}_{\mu\nu}\;,
\end{equation}
\begin{equation}
J_{\mu\nu}\equiv R_{\mu\nu}-\frac{1}{2}g_{\mu\nu}R\;,\quad
J\equiv g^{\mu\nu}J_{\mu\nu}\;.
\end{equation}
The currents are conserved:
\begin{equation}
\nabla^\mu{\hat J}_\mu=0\;,\quad\;\;\nabla^\mu J_{\mu\nu}=0\;.
\end{equation}
If all the curvatures vanish, the background is trivial. The
effective action is a functional of the curvatures (3.15).
\subsubsection{Redundancy of The Curvatures}
The effective action is a nonlocal functional of the curvatures,
and this fact conditions a certain simplification.

Since the commutator curvature is a commutator, it satisfies
the Jacobi identity, and so does the Riemann curvature:
\begin{equation}
\nabla_\gamma{\hat{\cal R}}_{\mu\nu}+
\nabla_\nu{\hat{\cal R}}_{\gamma\mu}+
\nabla_\mu{\hat{\cal R}}_{\nu\gamma}=0\;,
\end{equation}
\begin{equation}
\nabla_\gamma R_{\alpha\beta\mu\nu}+
\nabla_\nu R_{\alpha\beta\gamma\mu}+
\nabla_\mu R_{\alpha\beta\nu\gamma}=0\;.
\end{equation}
Act on these identities with $\nabla^\gamma$. In the first term, 
the operator $\Box$ forms, and in the remaining terms commute
the covariant derivatives. The commutator brings an extra power
of the curvature. The equations obtained
\begin{equation}
\Box{\hat{\cal R}}_{\mu\nu}+O(\Re^2)=
2\nabla_{[\nu}{\hat J}_{\mu]}\;,
\end{equation}
\begin{equation}
\Box R_{\alpha\beta\mu\nu}+O(\Re^2)=
4\nabla_{[\mu}\nabla_{\langle\alpha}\left(
J_{\nu]\beta\rangle}-\frac{1}{2}g_{\nu]\beta\rangle}J\right)
\end{equation}
hold identically and have the form of inhomogeneous
wave equations, the role of inhomogeneity being played by
the currents. In (3.21), (3.22), the brackets of both types $[\,]$ and 
$\langle\,\rangle$ denote the antisymmetrization in the respective 
indices.

The equations (3.21) and (3.22) are nonlinear but they can be solved
by iteration. The result is that the commutator and Riemann
curvatures get expressed in a nonlocal fashion through their
currents and an arbitrary solution of the homogeneous wave
equation
\begin{equation}
\Box{\hat{\cal R}}_{\mu\nu}^{\mbox{\small wave}}=0\;,\quad\;\;
\Box R_{\alpha\beta\mu\nu}^{\mbox{\small wave}}=0\;.
\end{equation}
If the metric is Lorentzian, this solution is fixed by initial
data which can be given in the remote past. It follows
that the commutator and Riemann curvatures are specified by
giving an incoming wave and the current $J$.
This fact underlies the Maxwell and Einstein equations. They
fix the currents $J$.
Adding initial conditions to these equations specifies the
connection and metric.

In the present case, since the metric is
Euclidean, there are no wave solutions:
\begin{equation}
{\hat{\cal R}}_{\mu\nu}^{\mbox{\small wave}}=0\;,\quad\;\;
R_{\alpha\beta\mu\nu}^{\mbox{\small wave}}=0\;,
\end{equation}
and the Green's function $1/\Box$ is unique. Therefore, the
commutator and Riemann curvatures are expressed entirely
through their currents:
\begin{equation}
{\hat{\cal R}}_{\mu\nu}=
\frac{1}{\Box}2\nabla_{[\nu}{\hat J}_{\mu]}+O(J^2)\;,
\end{equation}
\begin{equation}
R_{\alpha\beta\mu\nu}=
\frac{1}{\Box}4\nabla_{[\mu}\nabla_{\langle\alpha}\left(
J_{\nu]\beta\rangle}-\frac{1}{2}g_{\nu]\beta\rangle}J\right)
+O(J^2)\;.
\end{equation}

Thus, the curvatures are redundant because there are no waves 
in Euclidean theory. Owing to this fact, the set of field 
strengths (3.15) reduces to
\begin{equation}
\left(\,J_{\mu\nu}\,,\;
{\hat J}_\mu\,,\; {\hat P}\,\right)\;,
\end{equation}
and the effective action is a functional of the reduced set.
\subsubsection{The Axiomatic Effective Action}
To what class of functionals does the effective action belong?
One can say in advance that this should be a functional
analytic in the curvature. Indeed, the first variational
derivative of the effective action
taken at the trivial background should vanish
because, in the absence of an external source, the relative
vacuum becomes the absolute vacuum. The trivial background
should solve the mean-field equations in the absolute vacuum.
Higher-order variational derivatives taken at the trivial
background determine the correlation functions in the
absolute vacuum. They may not vanish but neither should they
blow up.

The analyticity suggests that the effective action can be built
as a sum of nonlocal invariants of $N$-th order in the curvature:
\begin{equation}
\Gamma=\sum_N\Gamma_N\;,\quad\;\;\Gamma_N=O[\Re^N]\;.
\end{equation}
Nonlocal invariant is, however, an uncertain concept.
Even local invariant of $N$-th
order in the curvature is a concept that needs to be refined
but this is easy to do. The most general local monomial that can be built
out of the available quantities yields an invariant of the form
\begin{equation}
\int dx\,g^{1/2}\underbrace{(\nabla_1{\scriptstyle\ldots}\nabla_1)
(\nabla_2{\scriptstyle\ldots}\nabla_2)\ldots}_{\displaystyle k}
\Re_1\Re_2\ldots\Re_N+O[\Re^{N+1}]\;.
\end{equation}
This monomial is a product of $N$ curvatures and $k$ covariant derivatives,
all indices being contracted by the metric. In
(3.29), the labels $1,2,\ldots$ point out which derivative acts
on which curvature but all the curvatures are at the same point,
and the total number of derivatives is finite. Of course, the
curvature sits also in the covariant derivatives and in the metric
that contracts the indices. Therefore, the $N$-th order invariant can
only be defined up to terms $O[\Re^{N+1}]$. In particular, the covariant
derivatives in (3.29) can be commuted freely because the contribution
of a commutator is already $O[\Re^{N+1}]$.

One may now consider a class of nonlocal invariants that can formally
be represented as infinite series of local invariants:
\begin{equation}
\Gamma_N=\int dx\,g^{1/2}\sum^\infty_{k=0}c_k
\underbrace{(\nabla_1{\scriptstyle\ldots}\nabla_1)
(\nabla_2{\scriptstyle\ldots}\nabla_2)\ldots}_{\displaystyle k}
\Re_1\Re_2\ldots\Re_N+O[\Re^{N+1}]\;.
\end{equation}
Here $c_k$ are some dimensional constants. It can be seen that this
is the needed class\footnote{To see it, consider any diagram
with massive propagators and expand it formally in the inverse mass.
The method that accomplishes this expansion is known as the
Schwinger--DeWitt technique.}. The number of curvatures in (3.30)
is $N$ but the number of derivatives is unlimited. Only a finite
number of derivatives can contract with the curvatures. The
remaining ones can only contract among themselves. If two
derivatives acting on the same curvature contract, they make
a $\Box$ operator acting on this curvature:
\begin{equation}
\nabla_1{}^2=\Box_1\;,\quad\nabla_2{}^2=\Box_2\;,\;\;\ldots\;.
\end{equation}
If two derivatives acting on different curvatures contract,
the contraction can again be written in terms of the $\Box$ operators:
\begin{eqnarray}
2\nabla_1\nabla_2&=&(\nabla_1+\nabla_2)^2-\nabla_1{}^2
-\nabla_2{}^2\nonumber\\
&=&\Box_{1+2}-\Box_1-\Box_2
\end{eqnarray}
but there appears a $\Box$ operator acting on 
the product of two curvatures:
\begin{equation}
\mathop{\Box_{1+2}}\Re_1\Re_2\Re_3\ldots=
\Box\left(\Re\Re\right)\,\Re_3\ldots\;.
\end{equation}
As a result, (3.30) takes the form
\begin{eqnarray}
\Gamma_N=\int dx\,g^{1/2}\left(\sum^\infty_{k_1,k_2,\cdots=0}c_k
(\Box_1)^{k_1}(\Box_2)^{k_2}(\Box_{1+2})^{k_3}\ldots\right)
\hphantom{a\Gamma+O[\Re^{N+1}]}{}
\nonumber\\
{}\times\underbrace{\Bigl(\nabla{\scriptstyle\ldots}\Re_1
\nabla{\scriptstyle\ldots}\Re_2\ldots
\nabla{\scriptstyle\ldots}\Re_N\Bigr)}_{\mbox{\small contraction}}
+O[\Re^{N+1}]\;.\nonumber\\
\end{eqnarray}
There remains an infinite series in the $\Box$ variables, and these 
variables themselves are operators acting on the curvatures in a given
contraction. The remaining series is some function of the $\Box$
variables:
\begin{equation}
\Gamma_N=\int dx\,g^{1/2}
F\left(\Box_1,\Box_2,\Box_{1+2},\ldots\right)
\underbrace{\Bigl(\nabla{\scriptstyle\ldots}\Re_1
\nabla{\scriptstyle\ldots}\Re_2\ldots
\nabla{\scriptstyle\ldots}\Re_N\Bigr)}_{\mbox{\small contraction}}
+O[\Re^{N+1}]\;.
\end{equation}
This is the general form of a nonlocal invariant of $N$-th order
in the curvature. The function $F$ is a formfactor.

There is, in addition, the identity
\begin{equation}
\nabla_1+\nabla_2+\ldots+\nabla_N=0
\end{equation}
which reduces the number of variables in the function $F$. The sum
in (3.36) is a derivative acting on the product of all curvatures,
i.e., a total derivative. Total derivatives vanish because the
curvatures may be considered having compact supports. Thus invariants
of first order in the curvature can only be local because any
derivative is a total derivative. Therefore, the first-order
formfactors are constants:
\begin{equation}
N=1\;:\qquad F={}\mbox{const.}
\end{equation}
At the second order, all formfactors are functions of only
one argument because the remaining arguments can be eliminated by
integration by parts:
\begin{equation}
N=2\;:\qquad F=F(\Box_1)\;,
\end{equation}
$$
\Box_2=\Box_1\;,\quad\Box_{1+2}=0\;.
$$
At the third order, all formfactors are functions of three
individual $\Box$'s because the $\Box$'s acting on pairs
can be eliminated:
\begin{equation}
N=3\;:\qquad F=F(\Box_1,\Box_2,\Box_3)\;,
\end{equation}
$$
\Box_{1+2}=\Box_3\;,\quad\Box_{1+3}=\Box_2\;,\quad
\Box_{2+3}=\Box_1\;.
$$
The $\Box$'s acting on pairs appear beginning with the fourth order 
in the curvature and are parameters of the on-shell scattering 
amplitudes.

Nonlocal invariants of a given order make a linear space in which
all possible contractions of $N$ curvatures and their derivatives
make a basis, and the formfactors play the role of coefficients
of the linear combining. The basis can be built by listing all
independent contractions. The effective action is an expansion
in this basis with certain coefficients--formfactors:
\begin{equation}
\Gamma=\Gamma_{\rm I}+\Gamma_{\rm II}+\Gamma_{\rm III}+
\ldots\;,
\end{equation}
\begin{equation}
\Gamma_{\rm I}=\int dx\,g^{1/2}
\Bigl[\,c_1R+c_2\mathop{\rm tr}{\hat P}\,\Bigr]\;,
\end{equation}
\begin{eqnarray}
\Gamma_{\rm II}=\int dx\,g^{1/2}\mathop{\rm tr}\,
\Bigl[\,R_{\mu\nu}&F_1(\Box)&R^{\mu\nu}\nonumber\\
{}+R&F_2(\Box)&R\nonumber\\
{}+{\hat P}&F_3(\Box)&R\nonumber\\
{}+{\hat P}&F_4(\Box)&{\hat P}\nonumber\\
{}+{\hat{\cal R}}_{\mu\nu}&F_5(\Box)&{\hat{\cal R}}^{\mu\nu}\,\Bigr]\;,
\end{eqnarray}
\begin{eqnarray}
\Gamma_{\rm III}=\int dx\,g^{1/2}\mathop{\rm tr}
&\Bigl[&
F_1(\Box_1,\Box_2,\Box_3)\,{\hat P}_1{\hat P}_2{\hat P}_3
\nonumber\\
&+&F_2(\Box_1,\Box_2,\Box_3)\,
{\hat{\cal R}}_1{}^\mu{}_\alpha
{\hat{\cal R}}_2{}^\alpha{}_\beta
{\hat{\cal R}}_3{}^\beta{}_\mu\nonumber\\
&+&\cdots\nonumber\\
&+&F_{29}(\Box_1,\Box_2,\Box_3)\,
\nabla_\lambda\nabla_\sigma R_1^{\alpha\beta}
\nabla_\alpha\nabla_\beta R_2^{\mu\nu}
\nabla_\mu\nabla_\nu R_3^{\lambda\sigma}\,\Bigr]\;.\nonumber\\
\end{eqnarray}
In the first-order action (3.41), there are 2 basis contractions:
the Ricci scalar and the trace of the matrix potential, and
the formfactors are constants. In the second-order action,
there are 5 independent contractions listed in (3.42). In the
third-order action, there are 29 basis contractions, examples
of which are given in (3.43). Here I shall stop because, for
the problems of interest, the third order is sufficient.
The reason for that will be explained in the next lecture.

In the expressions above, the basis invariants are written in
terms of the curvatures but they can be rewritten in terms
of the conserved currents. Note also that the operator arguments
of the third-order formfactors $F$ commute because they act on
different objects. Since the arguments commute, the functions $F$
themselves are ordinary functions of three variables.

Thus, even before any calculation, we have an ansatz for the 
effective action, with unknown formfactors.
We need them in the spectral forms
\begin{equation}
F_{\rm k}(\Box)=\int\limits_0^\infty dm^2\,
\frac{\rho_{\rm k}(m^2)}{m^2-\Box}
+\mbox{ a polynomial in }\Box\;,
\end{equation}
\begin{equation}
F_{\rm k}(\Box_1,\Box_2,\Box_3)=\int\limits_0^\infty 
dm_1^2dm_2^2dm_3^2\,
\frac{\rho_{\rm k}(m_1^2,m_2^2,m_3^2)}{(m_1^2-\Box_1)
(m_2^2-\Box_2)(m_3^2-\Box_3)}\;,
\end{equation}
and then we can proceed directly to the expectation-value equations.
Unknown are only the spectral weights. These are to be calculated
from the loop diagrams but there is an alternative approach.
One can look for the general limitations on the spectral weights stemming
from axiomatic theory. These limitations may be
sufficient to solve one's expectation-value problem. In this case,
the solution will prove to be independent of the details of the
quantum-field model and the approximations made in it. Moreover,
the effective action above does not refer even to quantum field theory.
It is an action for the observable field, and its implications
may be valid irrespective of the underlying fundamental theory.
Only certain axiomatic properties of the spectral weights may be
important. There is an example in which this approach has been
implemented \cite{53}.

Here, the axiomatic approach will not be considered. Let us see
how the effective action is calculated from loops.
\subsubsection{Heat Kernel}
Consider any diagram in the effective action
\begin{equation}
\parbox{42pt}{
\begin{picture}(42,42)(-21,-21)
\put(0,10.5){\line(-1,0){17.195}}
\put(0,10.5){\line(1,0){17.195}}
\put(0,-10.5){\line(-1,0){17.195}}
\put(0,-10.5){\line(1,0){17.195}}
\put(0,0){\line(1,1){13.85}}
\put(0,0){\line(-1,-1){13.85}}
\put(0,0){\circle{42}}
\end{picture}
}\;\,,
\end{equation}
and, for every propagator, write
\begin{equation}
\parbox{28pt}{
\begin{picture}(28,3)
\put(0,1.5){\line(1,0){28}}
\end{picture}
}
{}=-\frac{1}{H}=\int\limits_0^\infty ds\,{\rm e}^{sH}\;.
\end{equation}
The kernel of the exponential operator
\begin{equation}
{\rm e}^{sH}\delta(x,y)\equiv {\hat K}(x,y|s)
\end{equation}
(and the operator itself) is called heat kernel, and the
parameter $s$ is often called proper time. Both names are
matters of history, and a matter of physics is the fact
that $H$ is negative definite. The matrix $P$ in (3.8)
may spoil the negativity but, since it is treated
perturbatively, as one of the curvatures, this does not matter.

Upon the insertion of (3.47), the diagram remains the same 
as before but with
the heat kernels in place of the propagators, and the integrations
over the proper times will be left for the last:
\begin{equation}
\parbox{42pt}{
\begin{picture}(42,42)(-21,-21)
\put(0,10.5){\line(-1,0){17.195}}
\put(0,10.5){\line(1,0){17.195}}
\put(0,-10.5){\line(-1,0){17.195}}
\put(0,-10.5){\line(1,0){17.195}}
\put(0,0){\line(1,1){13.85}}
\put(0,0){\line(-1,-1){13.85}}
\put(0,0){\circle{42}}
\end{picture}
}
{}=\int\limits_0^\infty ds_1\ldots
\int\limits_0^\infty ds_n
\;\,
\parbox{42pt}{
\begin{picture}(42,42)(-21,-21)
\put(0,10.5){\line(-1,0){17.195}}
\put(0,10.5){\line(1,0){17.195}}
\put(0,-10.5){\line(-1,0){17.195}}
\put(0,-10.5){\line(1,0){17.195}}
\put(0,0){\line(1,1){13.85}}
\put(0,0){\line(-1,-1){13.85}}
\put(0,0){\circle{42}}
\put(-1,12.5){\llap{$s_1$}}
\put(8,-8.5){$s_n$}
\put(3,0){$\scriptstyle\ldots$}
\end{picture}
}\;\,.
\end{equation}
The one-loop effective action is the functional trace
of the heat kernel, integrated over $s$:
\begin{equation}
\parbox{20pt}{
\begin{picture}(20,20)
\put(10,10){\circle{20}}
\end{picture}
}
{}=\frac{1}{2}\ln\det\frac{1}{H}=\frac{1}{2}\int_0^\infty
\frac{ds}{s}\int dx\,\mathop{\rm tr}{\hat K}(x,x|s)\;.
\end{equation}
Thus, one is left with diagrams with the heat kernels.
It will be seen in a moment why this is better.

The expansion rule for the exponential operator has already been
considered in (1.27). There remains to be
presented the lowest-order approximation for the heat kernel:
\begin{equation}
{\hat K}(x,y|s)=\frac{1}{(4\pi s)^{D/2}}
\left({\rm e}^{-\sigma(x,y)/2s}{\hat a}(x,y)+O[\Re]\right)\;,
\end{equation}
\begin{equation}
D={}\mbox{dimension of the base manifold.}
\end{equation}
At the lowest order in the curvature, the potential $P$ does
not affect this expression but the metric and connection do.
As mentioned above, covariant expansions cannot be rigid.
In (3.51):
\begin{equation}
2\sigma(x,y)=(\mbox{geodetic distance between $x$ and $y$})^2
\end{equation}
in the metric entering the operator $H$. The connection entering
the operator $H$ defines a parallel transport along a line.
Parallel transport is a linear mapping, so there exists
a propagator of parallel transport (the matrix that accomplishes
this mapping). In (3.51):
\begin{equation}
{\hat a}(x,y)={}\parbox[t]{8cm}{propagator of the parallel transport
from $y$ to $x$\\along the geodesic connecting $y$ and $x$.}
\end{equation}
The geodesic comes from the metric, and the parallel transport
from the connection.

The two-point functions (3.53) and (3.54) are the main elements
of the Schwinger--DeWitt technique mentioned above and the basic
building blocks for all Green's functions: of the hyperbolic
operator $H$, and of the elliptic operator $H$, and the heat kernel.
What is special about the heat kernel? Special is the fact that,
as seen from expression (3.51), the heat kernel is finite at the
coincident points. Green's functions of the hyperbolic and
elliptic operators are singular, and this is normal. Abnormal
is the fact that in the loop diagrams they appear at the coincident
points. Finiteness of the heat kernel at the coincident points is
a bonus owing to which all diagrams with the heat kernels are
finite.

The divergences of the loop diagrams reappear in the proper-time
integrals in (3.49). These integrals diverge at the lower limits.
At this stage, one more advantage of the heat kernel comes into
effect. Namely, the manifold dimension $D$ enters only the
overall factor in (3.51). Apart from this factor, the expansion
of the heat kernel in the curvature does not contain $D$ explicitly.
Therefore, loops with the heat kernels are calculated once for
all dimensions, and then the knowledge of the analytic dependence
on $D$ enables one to apply the dimensional regularization to
the proper-time integrals. One integrates by parts in $s$ keeping
$\mathop{\rm Re}D<4$ and next goes over to the limit $D\to 4$.
For example,
\begin{equation}
\int\limits_0^\infty\frac{ds}{s^{D/2-1}}f(s)=\frac{1}{2-D/2}f(0)
-\int\limits_0^\infty ds\,\ln s \frac{df(s)}{ds}
+O\left(2-D/2\right)\;.
\end{equation}
The dimensional regularization annihilates all power divergences.
Only the logarithmic divergences survive and take the form of
poles in dimension. These poles affect only the polynomial terms
in the spectral representations of the formfactors. They appear in 
the coefficients of the polynomials, thereby making these coefficients
indefinite. As a consequence, the local terms of
the effective action will have indefinite coefficients.
I shall come back to this issue.

After the substitution of the heat kernels for the propagators,
the calculation of loops becomes an entertaining geometrical
exercise.
\subsubsection{Loops and Geometry}
The heat kernel involves $\sigma$ and ${\hat a}$.
The derivative of $\sigma$
\begin{equation}
\nabla^\mu \sigma(x,y)\equiv\sigma^\mu(x,y)
\;\;\qquad
\parbox{90pt}{
\begin{picture}(90,25.7)(-31.5,-10.7)
{\thicklines
\qbezier(-30,0)(0,20)(15,10.04)
}
\put(15,10.04){\vector(3,-2){11.8}}
\put(-30,0){\circle*{3}}
\put(15,10.04){\circle*{3}}
\put(-25.5,-5.7){$y$}
\put(15,14.54){\llap{$x$}}
\put(29.8,1.8){$\sigma^\mu(x,y)$}
\end{picture}
}
\end{equation}
is the vector tangent to the geodesic connecting $y$ and $x$,
directed outwards, and normalized to the geodetic distance
between $y$ and $x$:
\begin{equation}
g_{\mu\nu}\sigma^\mu\sigma^\nu=2\sigma\;,\quad\;\;
\sigma^\mu\Bigl|_{x=y}=0\;,\quad\;
\det\nabla^\nu\sigma^\mu\Bigl|_{x=y}\ne 0\;.
\end{equation}
The normalization condition is a closed equation for $\sigma$
which together with the conditions at the coincident points
can serve as the definition of $\sigma$. The defining equation
for ${\hat a}$ together with the condition at the coincident points is
\begin{equation}
\sigma^\mu\nabla_\mu{\hat a}(x,y)=0\;,\quad\;\;
{\hat a}\Bigl|_{x=y}={\hat 1}\;.
\end{equation}
The determinant
\begin{equation}
\det\Bigl(\nabla^x_\mu\nabla^y_\nu\sigma(x,y)\Bigr)=
g^{1/2}(x)g^{1/2}(y)\Delta (x,y)
\end{equation}
is known as the Van Vleck--Morette determinant. It is responsible,
in particular, for a caustic of the geodesics
emanating from $x$ or $y$.

The vector $\sigma^\mu$ can be used to expand any function in
a covariant Taylor series. For a scalar, this series is of the form
\begin{equation}
f(y)=\sum_{n=0}^\infty\frac{(-1)^n}{n!}\sigma^{\mu_1}\ldots
\sigma^{\mu_n}\nabla_{\mu_1}\ldots\nabla_{\mu_n}f(x)\;.
\end{equation}
If $f$ is not a scalar, it should at first be parallel
transported from $y$ to $x$:
\begin{equation}
f(y)={\hat a}(y,x)\sum_{n=0}^\infty\frac{(-1)^n}{n!}
\sigma^{\mu_1}\ldots
\sigma^{\mu_n}\nabla_{\mu_1}\ldots\nabla_{\mu_n}f(x)\;.
\end{equation}
The covariant Taylor expansion is a regrouping of the ordinary
Taylor expansion. Whatever the connection is, it cancels in
this series. The series can formally be written in the
exponential form
\begin{equation}
f(y)={\hat a}(y,x)\exp\left(-\sigma^\mu\nabla_\mu\right)f(x)
\end{equation}
which will be of use below. Two-point functions expanded
in this way get expressed through their covariant derivatives
at the coincident points. Thus
\begin{equation}
\Delta (x,y)=1+\frac{1}{6}R_{\mu\nu}\sigma^\mu\sigma^\nu+\ldots\;.
\end{equation}

A loop always involves the ring of ${\hat a}$'s
\begin{equation}
{\hat a}(x,x_1){\hat a}(x_1,x_2)\ldots{\hat a}(x_n,x)
\;,\;\qquad
\parbox{35pt}{
\begin{picture}(35,28)(-14,-15)
\put(-7,14){\line(1,0){14}}
\put(-7,-14){\line(1,0){14}}
\put(-14,0){\line(1,2){7}}
\put(-14,0){\line(1,-2){7}}
\put(14,0){\line(-1,2){7}}
\put(14,0){\line(-1,-2){7}}
\qbezier(14,14)(24,0)(14,-14)
\put(14,-14){\vector(-4,-3){3}}
\end{picture}
}
\end{equation}
i.e., the parallel transport around a geodetic polygon.
The ring of two ${\hat a}$'s is the parallel transport 
there and back along the same path. Therefore,
\begin{equation}
{\hat a}(x,x_1){\hat a}(x_1,x)\equiv {\hat 1}\;.
\end{equation}
The ring of three ${\hat a}$'s is the parallel transport around
the geodetic triangle. It involves the commutator curvature,
and the curvature terms can be calculated:
\begin{equation}
{\hat a}(x,x_1){\hat a}(x_1,x_2){\hat a}(x_2,x)={\hat 1}+
\frac{1}{2}{\hat{\cal R}}_{\alpha\beta}
\sigma_1{}^\alpha\sigma_2{}^\beta+\ldots\;,
\end{equation}
\begin{equation}
\parbox{77pt}{
\begin{picture}(77,54)(-32,-7)
{\thicklines
\put(0,20){\line(3,2){30}}
\put(0,20){\line(3,-2){30}}
\put(30,0){\line(0,1){40}}
}
\put(0,20){\vector(-3,2){12}}
\put(0,20){\vector(-3,-2){12}}
\put(6,18.55){$x$}
\put(33,39){$x_1$}
\put(33,-2){$x_2$}
\put(-12,28){\llap{$\sigma_2{}^\mu$}}
\put(-12,12){\llap{$\sigma_1{}^\mu$}}
\end{picture}
}\;\,.
\end{equation}
This is sufficient 
because any polygon can be broken into triangles:
\begin{equation}
\parbox{28pt}{
\begin{picture}(28,28)(-14,-14)
{\thicklines
\put(-7,14){\line(1,0){14}}
\put(-7,-14){\line(1,0){14}}
\put(-14,0){\line(1,2){7}}
\put(-14,0){\line(1,-2){7}}
\put(14,0){\line(-1,2){7}}
\put(14,0){\line(-1,-2){7}}
}
\put(14,0){\line(-3,2){21}}
\put(14,0){\line(-3,-2){21}}
\put(14,0){\line(-1,0){28}}
\put(2.5,3){\vector(-1,0){9.5}}
\put(-7,-3){\vector(1,0){9.5}}
\end{picture}
}\;\,.
\end{equation}
Solution of the geodetic triangle is also involved.
In the notation of (3.67),
\begin{equation}
\Bigl(\sigma^\mu(x_1,x_2)\Bigr)^2=\sigma_1{}^2+\sigma_2{}^2
-2\sigma_1\sigma_2-\frac{1}{3}R_{\mu\alpha\nu\beta}
\sigma_1{}^\mu\sigma_1{}^\nu\sigma_2{}^\alpha\sigma_2{}^\beta
+\ldots\;.
\end{equation}
Here the first two terms make the Pythagorean theorem,
the third term accounts for the angle not being the right
angle, and the terms with the Riemann curvature can be calculated.

The above is to give a flavour of what loops imply.
\subsubsection{Calculation of Loops}
The heat kernel calculates loops with a remarkable elegance.
As an example, consider the contribution of the second order
in the curvature to the effective action. The respective
one-loop diagram contains two curvatures $\Re$ and two
heat kernels with the proper times $s_1$ and $s_2$:
$$
\parbox{80pt}{
\begin{picture}(80,40)(-20,0)
\qbezier(0,20)(20,40)(40,20)
\qbezier(0,20)(20,0)(40,20)
\put(-7.30,20){\circle*{20}}
\put(47.30,20){\circle*{20}}
\put(-7.30,2.86){\llap{$\Re$}}
\put(47.30,2.86){$\Re$}
\put(20,33.5){$s_1$}
\put(20,3.1){$s_2$}
\end{picture}
}
{}+O[\Re^3]
$$
\begin{equation}
{}=\int dx\,g^{1/2}\int dy\,g^{1/2}\,\Re(x)
{\hat K}(x,y|s_1){\hat K}(x,y|s_2)\Re(y)+O[\Re^3]\;.
\end{equation}
Suppose that the calculation only needs to be done with
accuracy $O[\Re^3]$. Then one can insert in (3.70) the
lowest-order approximation for the heat kernels.
In this approximation, the rings of ${\hat a}$'s collapse
to ${\hat 1}$, and the remaining ${\hat a}$'s always
transport the $\Re$'s to the same point arranging their
complete contraction. With the ${\hat a}$'s and the numerical
coefficients omitted, the diagram (3.70) is of the form
\begin{eqnarray}
\frac{1}{s_1{}^{D/2}}\frac{1}{s_2{}^{D/2}}
\int dx\,g^{1/2}\int dy\,g^{1/2}
\hphantom{
\quad\left(-\frac{\sigma(x,y)}{2s_1}\right)
\exp\left(-\frac{\sigma(x,y)}{2s_2}\right)
}\nonumber\\
{}\times\Re(x)\exp\left(-\frac{\sigma(x,y)}{2s_1}\right)
\exp\left(-\frac{\sigma(x,y)}{2s_2}\right)
\Re(y)\;.
\end{eqnarray}
But the exponents here simply add, and the two heat kernels
turn into one with a complicated proper-time argument:
$$
\frac{1}{(s_1s_2)^{D/2}}\int dx\,g^{1/2}\int dy\,g^{1/2}\,\Re(x)
\exp\left(-\frac{s_1+s_2}{2s_1s_2}\sigma(x,y)\right)
\Re(y)\hphantom{{}={}={}={}}
$$
\begin{equation}
{}=\frac{1}{(s_1+s_2)^{D/2}}\int dx\,g^{1/2}\int dy\,g^{1/2}\,\Re(x)
K\left(x,y\Bigl|\frac{s_1s_2}{s_1+s_2}\right)
\Re(y)\;.
\end{equation}
One only needs to rewrite this heat kernel in the operator form:
\begin{equation}
\frac{1}{(s_1+s_2)^{D/2}}\int dx\,g^{1/2}\,\Re
\exp\left(\frac{s_1s_2}{s_1+s_2}\Box\right)
\Re(y)\;,
\end{equation}
and the loop is done. The proper-time integral
\begin{equation}
\int\limits_0^\infty ds_1
\int\limits_0^\infty ds_2\,
\frac{1}{(s_1+s_2)^{D/2}}
\exp\left(\frac{s_1s_2}{s_1+s_2}\Box\right)
=F(\Box)
\end{equation}
is the formfactor.

What has happened? The propagators in the loop glued together,
and the loop turned into a tree:
\begin{equation}
\parbox{200pt}{
\begin{picture}(80,40)(-20,0)
\qbezier(0,20)(20,40)(40,20)
\qbezier(0,20)(20,0)(40,20)
\put(-7.30,20){\circle*{20}}
\put(47.30,20){\circle*{20}}
\put(70,20){\vector(1,0){20}}
\put(100,0){\begin{picture}(80,40)(-20,0)
{\linethickness{3pt}
\put(0,20){\line(1,0){40}}
}
\put(-7.30,20){\circle*{20}}
\put(47.30,20){\circle*{20}}
\end{picture}}
\end{picture}
}\;\,.
\end{equation}
This is what means to do the loop. {\it It means to
turn it into a tree.} The role of the propagator in
the tree is played by the formfactor $F(\Box)$.

Consider now any multi-loop diagram with parallel propagators.
It turns into a tree
\begin{equation}
\parbox{200pt}{
\begin{picture}(80,40)(-20,0)
\qbezier(0,20)(20,40)(40,20)
\qbezier(0,20)(20,0)(40,20)
\qbezier(0,20)(20,25)(40,20)
\qbezier(0,20)(20,15)(40,20)
\qbezier(0,20)(20,50)(40,20)
\qbezier(0,20)(20,-10)(40,20)
\put(-7.30,20){\circle*{20}}
\put(47.30,20){\circle*{20}}
\put(70,20){\vector(1,0){20}}
\put(100,0){\begin{picture}(80,40)(-20,0)
{\linethickness{3pt}
\put(0,20){\line(1,0){40}}
}
\put(-7.30,20){\circle*{20}}
\put(47.30,20){\circle*{20}}
\end{picture}}
\end{picture}
}
\end{equation}
in a completely similar way. The inverse proper times add:
$$
\frac{1}{s_1}+\frac{1}{s_2}+\ldots=
\frac{1}{s_{\mbox{\scriptsize total}}}
$$
(the law of parallel conductors). There is nothing to do.

For more than two curvatures a more powerful method is used.
Consider the diagram
\begin{equation}
\parbox{80pt}{
\begin{picture}(80,70)(-40,-20)
\put(-20,0){\line(1,0){40}}
\put(0,30){\line(-2,-3){20}}
\put(0,30){\line(2,-3){20}}
\put(-26.53,-3.265){\circle*{20}}
\put(26.53,-3.265){\circle*{20}}
\put(0,37.30){\circle*{20}}
\put(-17,8){\llap{$y_1$}}
\put(17,8){$y_2$}
\put(6,27){$x$}
\end{picture}
}
{}+O[\Re^4]\;,
\end{equation}
and suppose again that it is needed only up to the
next order in the curvature. Then, with the ${\hat a}$'s
and the numerical coefficients omitted, it is of the form
\begin{eqnarray}
\frac{1}{s_1{}^{D/2}}\frac{1}{s_2{}^{D/2}}
\frac{1}{s_3{}^{D/2}}
\int dx\,g^{1/2}\int dy_1\,g^{1/2}\int dy_2\,g^{1/2}
\hphantom{
\Re(y_1)\Re(y_2)\Re\;{}
}\nonumber\\
{}\times\exp\left(-\frac{\sigma(x,y_1)}{2s_1}
-\frac{\sigma(x,y_2)}{2s_2}
-\frac{\sigma(y_1,y_2)}{2s_3}\right)\Re(x)\Re(y_1)\Re(y_2)\;.
\;{}
\end{eqnarray}
Choose one of the vertices, say $x$, to be the observation point
of the effective Lagrangian. One of the curvatures, $\Re(x)$,
is already there. Shift the remaining curvatures to $x$ using
the covariant Taylor series:
\begin{equation}
\Re(y_i)=\exp\left(-\sigma_i{}^\mu\nabla_\mu\right)\Re(x)\;,
\end{equation}
\begin{equation}
\sigma_i{}^\mu=\sigma^\mu(x,y_i)\;,
\quad i=1,2\,.
\end{equation}
Next, consider the geodetic triangle with the same vertices
as in the diagram. For the geodesics connecting $x$ with
$y_i$, write
\begin{equation}
2\sigma(x,y_i)=(\sigma_i)^2\;,
\end{equation}
and, for the geodesic between the $y$'s, use the Pythagorean
theorem:
\begin{equation}
2\sigma(y_1,y_2)=(\sigma_1)^2+(\sigma_2)^2
-2\sigma_1\sigma_2+O[\Re]\;.
\end{equation}
Finally, replace the integration variables:
\begin{equation}
y_1{}^\mu\to\sigma_1{}^\mu\;,\quad
y_2{}^\mu\to\sigma_2{}^\mu\;.
\end{equation}
The Jacobian
\begin{equation}
\left|\frac{\partial\sigma^\mu(x,y_i)}{\partial y_i{}^\nu}
\right|^{-1}=\frac{g^{1/2}(x)}{g^{1/2}(y_i)}\Delta^{-1}(x,y_i)
=\frac{g^{1/2}(x)}{g^{1/2}(y_i)}\left(1+O[\Re]\right)
\end{equation}
removes the measure $g^{1/2}$ from the integral in $y_i$ and
brings an extra $g^{1/2}$ to the integral in $x$. Expression
(3.78) takes the form
\begin{eqnarray}
\frac{1}{(s_1s_2s_3)^{D/2}}\int dx\, g^{1/2}\,
\left(g^{1/2}(x)\right)^2\int d\sigma_1d\sigma_2
\exp\left(-\frac{\sigma_1{}^2}{4s_1}
-\frac{\sigma_2{}^2}{4s_2}
\hphantom{(x)\quad{}}
\vphantom{\frac{\sigma_1{}^2+\sigma_2{}^2-2\sigma_1\sigma_2}{4s_3}}
\right.\nonumber\\
{}-\left.\frac{\sigma_1{}^2+\sigma_2{}^2-2\sigma_1\sigma_2}{4s_3}
-\sigma_1{}^\mu\nabla_\mu{}^1
-\sigma_2{}^\mu\nabla_\mu{}^2\right)
\Re(x)\Re_1(x)\Re_2(x)\;.
\quad{}
\end{eqnarray}
Here the labels $1,2$ on $\nabla_\mu$ and $\Re$ point out which
$\nabla_\mu$ acts on which $\Re$. The operators $\nabla_\mu$
figure as parameters in the integral, and, up to the next order
in $\Re$, they commute. Since the parameters commute, the integral
in $\sigma_1{}^\mu$, $\sigma_2{}^\mu$ is an ordinary Gaussian
integral. Do it. The extra factor $\left(g^{1/2}(x)\right)^2$
cancels, and the result is
\begin{equation}
B(s_1,s_2,s_3)\int dx\,g^{1/2}\exp\left(\sum_{i,k=1}^2 
b_{ik}(s_1,s_2,s_3)
\nabla_i\nabla_k\right)\Re(x)\Re_1(x)\Re_2(x)
\end{equation}
where $B(s_1,s_2,s_3)$ is some function of the proper times,
and the exponent is a quadratic form in $\nabla_1$, $\nabla_2$
with $s$-dependent coefficients. The loop is done. The integral
\begin{eqnarray}
\int\limits_0^\infty ds_1ds_2ds_3\,
B(s_1,s_2,s_3)\exp\left(\sum_{i,k=1}^2
b_{ik}(s_1,s_2,s_3)
\nabla_i\nabla_k\right)
\hphantom{\nabla_2)}\nonumber\\
{}=F(\nabla_1{}^2,\nabla_2{}^2,\nabla_1\nabla_2)
\end{eqnarray}
is the formfactor. Integration by parts in $x$ brings it
to the $\Box$ arguments:
\begin{equation}
F(\nabla_1{}^2,\nabla_2{}^2,\nabla_1\nabla_2)\to
F(\nabla_1{}^2,\nabla_2{}^2,\nabla^2)\;.
\end{equation}

The effect of the calculation above is again that the loop
is turned into a tree:
\begin{equation}
\parbox{200pt}{
\begin{picture}(80,70)(-40,-20)
\put(-20,0){\line(1,0){40}}
\put(0,30){\line(-2,-3){20}}
\put(0,30){\line(2,-3){20}}
\put(-26.53,-3.265){\circle*{20}}
\put(26.53,-3.265){\circle*{20}}
\put(0,37.30){\circle*{20}}
\put(50,15){\vector(1,0){20}}
\put(80,-20){\begin{picture}(80,70)(-40,-20)
{\linethickness{3pt}
\put(-20,0){\line(1,0){40}}
\put(0,0){\line(0,1){30}}
}
\put(-27.30,0){\circle*{20}}
\put(27.30,0){\circle*{20}}
\put(0,37.30){\circle*{20}}
\end{picture}}
\end{picture}
}\;\,.
\end{equation}
The vertex of the tree is the formfactor
$F(\nabla_1{}^2,\nabla_2{}^2,\nabla_3{}^2)$. 
This method applies to any diagram with the heat kernels. One
only needs to do Gaussian integrals, and the result is always
the exponential of a quadratic combination of $\nabla$'s.
The formfactor is a function of the products $\nabla_i\nabla_k$.
\subsubsection{The One-Loop Formfactors}
The result of the proper-time integrations depends essentially
on the dimension~$D$. For $D=4$, the one-loop formfactors in the
effective action (3.40) are as follows.

With one exception, all second-order formfactors are logs:
\begin{eqnarray}
F_1(\Box)&=&\frac{1}{60}\frac{1}{2(4\pi)^2}\ln(-\Box)
+\mbox{ const.}\;,\\
F_2(\Box)&=&-\frac{1}{180}\frac{1}{2(4\pi)^2}\ln(-\Box)
+\mbox{ const.}\;,\\
F_3(\Box)&=&\frac{1}{18}\frac{1}{2(4\pi)^2}\;,\\
F_4(\Box)&=&\frac{1}{2}\frac{1}{2(4\pi)^2}\ln(-\Box)
+\mbox{ const.}\;,\\
F_5(\Box)&=&\frac{1}{12}\frac{1}{2(4\pi)^2}\ln(-\Box)
+\mbox{ const.}
\end{eqnarray}
Since
\begin{equation}
-\ln(-\Box)=\int\limits_0^\infty dm^2\,\frac{1}{m^2-\Box}
+\mbox{ const.}\;,
\end{equation}
these expressions have the spectral forms (3.44) with definite
spectral weights and indefinite additive constants (polynomials
of the zeroth power). Respectively, the effective action contains
a set of local terms with unspecified coefficients:
\begin{eqnarray}
\Gamma&=&\frac{1}{2(4\pi)^2}\int dx\,g^{1/2}\Bigl(
c_1R+c_2\mathop{\rm tr}{\hat P}+c_3R_{\mu\nu}R^{\mu\nu}
+c_4R^2\nonumber\\
&&{}+c_5\mathop{\rm tr}({\hat P}{\hat P})
+c_6\mathop{\rm tr}({\hat{\cal R}}_{\mu\nu}
{\hat{\cal R}}^{\mu\nu})+\frac{1}{18}R\mathop{\rm tr}{\hat P}
+\mbox{ nonlocal terms}\Bigr)\;.\hphantom{\qquad{}}
\end{eqnarray}
The nonlocal terms are specified completely.

The third-order formfactors have no polynomial terms and
indefinite coefficients. The simplest third-order formfactor is
${F_1(\Box_1,\Box_2,\Box_3)}$ in (3.43). It has the spectral form
(3.45), and its spectral weight ${\rho_1(m_1^2,m_2^2,m_3^2)}$ is
obtained as follows. Consider a triangle of three spectral masses
\begin{center}
\parbox{190pt}{
\begin{picture}(190,37)(-20,-7)
\put(-20,0){\line(1,0){40}}
\put(0,30){\line(-2,-3){20}}
\put(0,30){\line(2,-3){20}}
\put(-13,13.2){\llap{$m_1$}}
\put(13,13.2){$m_2$}
\put(-4,-6.6){$m_3$}
\put(60,13.2){$A={}$area of the triangle.}
\end{picture}
}
\end{center}
It can be built only if every mass is smaller than the sum
of the two others. The spectral weight $\rho_1$ is zero if
the triangle cannot be built. Otherwise, it is proportional
to the inverse area of this triangle:
\begin{eqnarray}
\rho_1(m_1^2,m_2^2,m_3^2)=-\frac{1}{3}\frac{1}{2(4\pi)^2}
\frac{1}{4\pi A}
\hphantom{\theta(m_1+m_2-m_3)
\theta(m_1+m_3}\nonumber\\
{}\times\theta(m_1+m_2-m_3)
\theta(m_1+m_3-m_2)
\theta(m_2+m_3-m_1)\;.
\end{eqnarray}

The remaining 28 third-order formfactors are expressed through
$F_1$ and are tabulated \cite{36}. The tables
contain various integral representations of the formfactors,
and their asymptotics.

The loop of the minimal second-order operator with arbitrary
metric, connection, and potential is called standard loop
because every calculation with it is done once, and the
results can be tabulated. A calculation in any specific model
boils down to combining the standard loops and using the tables.
A number of recipes for the reduction to minimal operators
can be found in \cite{24}. Doing loops becomes a business similar
to doing integrals.

The fact that some coefficients in the effective action
remain unspecified is none of the tragedy. The effective
action is a phenomenological object intended for obtaining
the values of observables. The spectral weights are certain
phenomenological characteristics of the vacuum like the
permittivity of a medium. They are to be calculated from
a more fundamental microscopic theory. Some microscopic
theory of some level is incapable of specifying some of
the coefficients. So what?  Classical theory was capable
of even less, and, nevertheless, celestial mechanics has
been successfully worked up\footnote{Remarkably, without
a knowledge of string theory!}. The only important question
is whether the lack of knowledge affects the problems that
we want to solve. This will be cleared up in the next lecture.

}

%% file: lecture4.tex
\section[Lecture 4]{Vacuum Currents and The Effect of Particle Creation}
\label{sec:4}
{\renewcommand{\theequation}{4.\arabic{equation}}
\subsubsection{Vacuum Currents}
Consider quantum electrodynamics. In this case, $\f^a(x)$
is a set of the vector connection field and the electron--positron
field
\begin{equation}
\mbox{QED:}\;\;\quad \f^a=\Bigl({\cal A}_\mu,\psi\Bigr)\;.
\end{equation}
The commutator curvature is, up to a coefficient, 
the Maxwell tensor, and the
operator field equations are of the form
\begin{equation}
\nabla^\nu{\cal R}_{\nu\mu}({\hat{\cal A}})
+J_\mu({\hat\psi})
=-J_\mu^{\mbox{\scriptsize ext}}
\end{equation}
where $J_\mu({\hat\psi})$ is the operator electron--positron 
current, and $J_\mu^{\mbox{\scriptsize ext}}$ is an external
source. Averaging these equations over the in-vacuum state,
one obtains, according to the general derivation above,
the same terms but as functions of the mean field plus a set
of loops:
\begin{equation}
\nabla^\nu{\cal R}_{\nu\mu}(\langle{\cal A}\rangle)
+{}
\parbox{31.4pt}{
\begin{picture}(31.4,32)(0,-6)
\put(0,7){$J_\mu(\langle\psi\rangle)$}
\put(3,-6){\line(1,1){33}}
\end{picture}
}
{}+{}
\parbox{30pt}{
\begin{picture}(30,32)(0,-6)
{\thicklines
\put(20,10){\circle{20}}
\put(0,10){\line(1,0){10}}
\put(0,12){${\scriptstyle{\cal A}}$}
\put(19,22){\llap{${\scriptstyle{\cal A}}$}}
\put(18,-6){\llap{${\scriptstyle{\cal A}}$}}
}
\put(3,-6){\line(1,1){33}}
\end{picture}
}
{}+{}
\parbox{30pt}{
\begin{picture}(30,32)(0,-6)
{\thicklines
\put(20,10){\circle{20}}
\put(0,10){\line(1,0){10}}
\put(0,12){${\scriptstyle{\cal A}}$}
\put(19,22){\llap{$\psi$}}
\put(18,-6){\llap{${\scriptstyle{\cal A}}$}}
}
\put(3,-6){\line(1,1){33}}
\end{picture}
}
{}+{}
\parbox{30pt}{
\begin{picture}(30,32)(0,-6)
{\thicklines
\put(20,10){\circle{20}}
\put(0,10){\line(1,0){10}}
\put(0,12){${\scriptstyle{\cal A}}$}
\put(19,22){\llap{${\scriptstyle{\cal A}}$}}
\put(18,-6){\llap{$\psi$}}
}
\put(3,-6){\line(1,1){33}}
\end{picture}
}
{}+{}
\parbox{30pt}{
\begin{picture}(30,32)(0,-6)
\thicklines
\put(20,10){\circle{20}}
\put(0,10){\line(1,0){10}}
\put(0,12){${\scriptstyle{\cal A}}$}
\put(19,22){\llap{$\psi$}}
\put(18,-6){\llap{$\psi$}}
\end{picture}
}
{}=-J_\mu^{\mbox{\scriptsize ext}}\;.
\end{equation}
There is another such equation, for $\psi$, but, since
$\psi$ has no external source, its solution is
\begin{equation}
\langle\psi\rangle=0\;.
\end{equation}
Then, in (4.3), $J_\mu(\langle\psi\rangle)$ vanishes, and
the loops with the vertices $S_{{\cal A}{\cal A}{\textstyle\psi}}$
vanish. There are no such vertices in QED but, if there were,
as in gravidynamics, they would be proportional to
$\langle\psi\rangle$ and vanish by (4.4). The photon loop
also vanishes because neither there is a vertex 
$S_{{\cal A}{\cal A}{\cal A}}$ but this is already a specific
property of QED. Only the electron--positron loop survives.

The surviving loop is a function of $\langle{\cal A}\rangle$,
and, by derivation, is the electron--positron current
averaged over the in-vacuum:
\begin{equation}
\parbox{30pt}{
\begin{picture}(30,32)(0,-6)
\thicklines
\put(20,10){\circle{20}}
\put(0,10){\line(1,0){10}}
\put(0,12){${\scriptstyle{\cal A}}$}
\put(19,22){\llap{$\psi$}}
\put(18,-6){\llap{$\psi$}}
\end{picture}
}
{}=J_\mu^{\mbox{\scriptsize vac}}
(\langle{\cal A}\rangle)=
\langle\mbox{in vac}|
J_\mu({\hat\psi})|\mbox{in vac}\rangle\;.
\end{equation}
This is the vacuum current. According to (4.3),
the {\it observable} electromagnetic field satisfies
the Maxwell equations with an addition of the vacuum current:
\begin{equation}
\nabla^\nu{\cal R}_{\nu\mu}({\cal A})
=-J_\mu^{\mbox{\scriptsize vac}}({\cal A})
-J_\mu^{\mbox{\scriptsize ext}}\;.
\end{equation}
We obtain this current by varying the effective action and
next replacing the Euclidean resolvents with the retarded
resolvents:
\begin{equation}
J_\mu^{\mbox{\scriptsize vac}}({\cal A})=\left.
\frac{\delta\Gamma({\cal A})}{\delta{\cal A}^\mu}
\right|_{\Box\to\Box_{\mbox{\scriptsize ret}}}\;,
\end{equation}
\begin{equation}
\Gamma({\cal A})=\int dx\,g^{1/2}\Bigl[{\cal R}F(\Box){\cal R}
+F(\Box_1,\Box_2,\Box_3){\cal R}_1{\cal R}_2{\cal R}_3+\ldots
\Bigr]\;.
\end{equation}

It is completely similar if $\f^a(x)$ is a set of the metric field
and any matter fields
\begin{equation}
\mbox{GRAVITY:}\;\;\quad \f^a=\Bigl(g_{\mu\nu},\psi\Bigr)\;.
\end{equation}
The only difference is that the vertex $S_{ggg}$ is nonvanishing:
\begin{equation}
R_{\mu\nu}(\langle g\rangle)-\frac{1}{2}\langle
g_{\mu\nu}\rangle R(\langle g\rangle)
+{}
\parbox{30pt}{
\begin{picture}(30,32)(0,-6)
\thicklines
\put(20,10){\circle{20}}
\put(0,10){\line(1,0){10}}
\put(0,13.5){$g$}
\put(19,22){\llap{$\psi$}}
\put(18,-6){\llap{$\psi$}}
\end{picture}
}
{}+{}
\parbox{30pt}{
\begin{picture}(30,32)(0,-6)
\thicklines
\put(20,10){\circle{20}}
\put(0,10){\line(1,0){10}}
\put(0,13.5){$g$}
\put(19,23){\llap{$g$}}
\put(18,-5){\llap{$g$}}
\end{picture}
}
{}=8\pi T_{\mu\nu}^{\mbox{\scriptsize ext}}\;,
\end{equation}
\begin{equation}
\langle\psi\rangle=0\;,
\end{equation}
and it is assumed again that the matter fields have no sources.
Again, by derivation, the matter loop is the energy-momentum 
tensor of the field ${\hat\psi}$ averaged over the in-vacuum but
the vacuum current contains, in addition, the graviton loop:
\begin{equation}
T_{\mu\nu}^{\mbox{\scriptsize vac}}=
=\langle\mbox{in vac}|
T_{\mu\nu}({\hat\psi})|\mbox{in vac}\rangle
+\mbox{ the graviton loop.}
\end{equation}
The Einstein equations are replaced by the expectation-value
equations in the in-vacuum state:
\begin{equation}
R_{\mu\nu}-\frac{1}{2}g_{\mu\nu}R
=8\pi T_{\mu\nu}^{\mbox{\scriptsize vac}}(g)
+8\pi T_{\mu\nu}^{\mbox{\scriptsize ext}}\;.
\end{equation}

Since the gravitational field couples to everything,
the equation (4.10) should contain loops of all matter fields
in Nature.
The effective actions for all loops including the graviton loop
have the same structure:
\begin{equation}
T_{\mu\nu}^{\mbox{\scriptsize vac}}(g)=-\frac{2}{g^{1/2}}\left.
\frac{\delta\Gamma(g)}{\delta g^{\mu\nu}}
\right|_{\Box\to\Box_{\mbox{\scriptsize ret}}}\;,
\end{equation}
\begin{equation}
\Gamma(g)=\int dx\,g^{1/2}\Bigl[R_{..}F(\Box)R_{..}
+F(\Box_1,\Box_2,\Box_3)R_{1..}R_{2..}R_{3..}+\ldots
\Bigr]\;.
\end{equation}
Only the coefficients of the formfactors are different.
To have the correct coefficients, one would need to know
the full spectrum of particles. Therefore, in the case of
gravity, the axiomatic approach is most suitable.

Now recall that the curvatures are redundant, and the effective
action is in fact a functional of the conserved currents
(3.16) and (3.17). Owing to this fact, the expectation-value
equations (4.6) and (4.13) close with respect to these currents:
\begin{equation}
\Bigl(\nabla^\nu{\cal R}_{\nu\mu}\Bigr)+
f(\Box_{\mbox{\scriptsize ret}})
\Bigl(\nabla^\nu{\cal R}_{\nu\mu}\Bigr)+
O\Bigl(\nabla^\nu{\cal R}_{\nu\mu}\Bigr)^2
=-J_\mu^{\mbox{\scriptsize ext}}\;,
\end{equation}
\begin{eqnarray}
\Bigl(R_{\mu\nu}-\frac{1}{2}g_{\mu\nu}R\Bigr)+
f_1(\Box_{\mbox{\scriptsize ret}})
\Bigl(R_{\mu\nu}-\frac{1}{2}g_{\mu\nu}R\Bigr)
\hphantom{+f_2(\Box_{\mbox{\scriptsize ret}})
(-g_{\mu\nu}\Box)+{}+{}
}\nonumber\\
{}+f_2(\Box_{\mbox{\scriptsize ret}})
(\nabla_\mu\nabla_\nu-g_{\mu\nu}\Box)R
+O\Bigl(R_{\mu\nu}-\frac{1}{2}g_{\mu\nu}R\Bigr)^2
=8\pi T_{\mu\nu}^{\mbox{\scriptsize ext}}\;.\quad{}
\end{eqnarray}
Of course, with respect to the mean fields, these equations
are closed from the outset but, at an intermediate stage,
they are closed with respect to the Maxwell and Einstein
currents. When solved with respect to these currents, they
become literally the Maxwell and Einstein equations with
some external sources but {\it not} the original ones.
To make this clear, use the fact that the vacuum terms
are proportional to the Planck constant and solve the
equations by iteration:
\begin{equation}
\nabla^\nu{\cal R}_{\nu\mu}=
=-J_\mu^{\mbox{\scriptsize ext}}
+f(\Box_{\mbox{\scriptsize ret}})
J_\mu^{\mbox{\scriptsize ext}}
+O\left(J_\mu^{\mbox{\scriptsize ext}}\right)^2\;,
\end{equation}
\begin{eqnarray}
R_{\mu\nu}-\frac{1}{2}g_{\mu\nu}R
=8\pi T_{\mu\nu}^{\mbox{\scriptsize ext}}
-f_1(\Box_{\mbox{\scriptsize ret}})
8\pi T_{\mu\nu}^{\mbox{\scriptsize ext}}
\hphantom{+f_2(\Box_{\mbox{\scriptsize ret}})+{}
}\nonumber\\
{}+f_2(\Box_{\mbox{\scriptsize ret}})
(\nabla_\mu\nabla_\nu-g_{\mu\nu}\Box)
8\pi T^{\mbox{\scriptsize ext}}
+O\left(T_{\mu\nu}^{\mbox{\scriptsize ext}}\right)^2\;.\quad{}
\end{eqnarray}
These are the Maxwell and Einstein equations with the original
sources propagated in a nonlocal and nonlinear manner.

There is an effect in these equations that drives the entire problem.
\subsubsection{Emission of Charges}
Consider again QED and suppose that the external source 
has a compact spatial support. This source is the current of
a set of electrically charged particles moving inside a
spacetime tube but, since the observable electromagnetic
field is the expectation value, only the total current in
(4.6) or (4.18) is observable:
\begin{equation}
J_\mu^{\mbox{\scriptsize tot}}
=J_\mu^{\mbox{\scriptsize ext}}
+J_\mu^{\mbox{\scriptsize vac}}({\cal A})\;.
\end{equation}
And the total current has a noncompact spatial support
because the vacuum contribution is nonlocal. One may
calculate the flux of charge through the support tube of
$J^{\mbox{\scriptsize ext}}$ and even through a wider
tube (see Fig. 3), and it will be nonvanishing:
\begin{equation}
e_{\cal T}(\Sigma_1)-
e_{\cal T}(\Sigma_2)=
\frac{1}{4\pi}\int\limits_{\Sigma_1}^{\Sigma_2}
J_\mu^{\mbox{\scriptsize vac}}\,d{\cal T}^\mu\ne 0\;.
\end{equation}
Here $e_{\cal T}(\Sigma)$ is the amount of the electric charge
contained inside the tube ${\cal T}$ at a given instant $\Sigma$.
The charge inside the tube is not conserved.
\input figthree.tex

If, when moving away from the support of
$J^{\mbox{\scriptsize ext}}$, the flux (4.21) falls off rapidly,
then its nonvanishing only means that the boundary of the
original source gets spread. Because of the creation of
virtual pairs, this boundary can never be located precisely.
The charges of the external source immersed in the quantum
vacuum are always annihilated and created again in a
slightly different place. There is no point to worry about.
Just step aside a little.

However, one may ask if there is a flux of charge through
an infinitely wide tube:
\begin{equation}
e(\Sigma_1)-
e(\Sigma_2)=
\frac{1}{4\pi}\int\limits_{\Sigma_1}^{\Sigma_2}
J_\mu^{\mbox{\scriptsize vac}}\,d{\cal T}^\mu
\Bigl|_{r\to\infty}\;.
\end{equation}
In this equation, $e(\Sigma)$ is the total amount of the
electric charge in the compact domain of space at a given
instant $\Sigma$. For (4.22) to be nonvanishing,
$J_\mu^{\mbox{\scriptsize vac}}$ should behave as
\begin{equation}
J_\mu^{\mbox{\scriptsize vac}}=O\left(\frac{1}{r^2}\right)\;,
\quad r\to\infty\;,
\end{equation}
\begin{equation}
r\propto\sqrt{\mbox{area of }{\bf{\cal S}}}
\end{equation}
where ${\bf{\cal S}}$ is the intersection of ${\cal T}$
with $\Sigma$ (Fig. 3). In this case, it would turn out
that the charge disappears, i.e., {\it our source is emitting 
charge}. But even this may not be a point of concern if the
current in (4.22) oscillates with time, and the oscillations sum
to zero for a sufficiently long period between $\Sigma_1$
and $\Sigma_2$. The expectation values have uncertainties,
and these oscillations are a quantum noise. Just do not
measure (4.22) too often.

However, one may ask if the charge emitted for the entire
history
\begin{equation}
e(-\infty)-
e(+\infty)=
\frac{1}{4\pi}\int\limits_{\Sigma\to -\infty}^{\Sigma\to +\infty}
J_\mu^{\mbox{\scriptsize vac}}\,d{\cal T}^\mu
\Bigl|_{r\to\infty}
\end{equation}
is nonvanishing. There will always be oscillations in the current
but they may sum not to zero. Since, as $r\to\infty$, all fields
fall off, there are, in this limit, the asymptotic Killing vectors
corresponding to all the symmetries of flat and empty spacetime.
Therefore, one may ask the same questions about the emission of
energy and any other charges. Thus the quantity
\begin{equation}
M(-\infty)-
M(+\infty)=
\int\limits_{\Sigma\to -\infty}^{\Sigma\to +\infty}
T_{\mu\nu}^{\mbox{\scriptsize vac}}\xi^\nu\,d{\cal T}^\mu
\Bigl|_{r\to\infty}
\end{equation}
with $\xi^\nu$ the asymptotic timelike Killing vector 
is the energy emitted by the source for the entire history.

If the total emitted charges are nonvanishing, then this is
the real effect, and then the question emerges: what are
the carriers of these charges? There should be some real
agents carrying them away. But the particles of the original
source stay in the tube. Besides them, there is only the
electron--positron field but it is in the in-vacuum state.
This means that, at least initially, there are neither
electrons nor positrons. There remains to be assumed
a miracle: that either the real electrons or the real
positrons -- depending on the sign of the emitted 
charge -- get created. Then they are created by pairs,
and, say, the created positron is emitted while the
created electron stays in the compact domain.

This crazy guess can be checked. We have two ways of
calculating the vacuum currents: through the effective
action and by a direct averaging of the operator currents
as in (4.5) and (4.12). Specifically, for the in-vacuum of
electrons and positrons we have
\begin{equation}
T_{\mu\nu}^{\mbox{\scriptsize vac}}=
=\langle\mbox{in vac}|
T_{\mu\nu}({\hat\psi})|\mbox{in vac}\rangle
\end{equation}
where $T_{\mu\nu}({\hat\psi})$ is the operator
energy-momentum tensor of the electron--positron field 
${\hat\psi}$. The equation for ${\hat\psi}$
\begin{equation}
\left(\lefteqn{\!\not}\partial+\mu-{\rm i}q\langle
\lefteqn{\,\not}{\cal A}\rangle
\right){\hat\psi}=0
\end{equation}
contains the electromagnetic field which in (4.27) figures
as an external field but is in fact the mean field solving
the expectation-value equations. We know that, in the past, 
all mean fields are static. In the future, they become
static again because, if the total emitted charges are finite,
then all the processes should die down. Thus, there are
two asymptotically static regions: in the past and in the
future. The carriers of the emitted charges should be
detectable in the future as particles with definite energies.
But then the state in which they are absent is the
out-vacuum whereas their quantum state is the in-vacuum.
{\it It may be the case that the in-vacuum contains the
out-particles.} This will be the case if, between the
static regions in the past and future, there is a region
where $\langle{\cal A}\rangle$ is nonstatic because then
the basis functions of the Fock modes that are the
eigenfunctions of the energy operator in the future and
the basis functions that are such in the past are different
solutions of the Dirac equation (4.28).

If we expand ${\hat\psi}$ in the basis solutions of the
out-particles, insert this expansion in (4.27), and then
insert (4.27) in (4.26), the result will be
\begin{equation}
M(-\infty)-M(+\infty)=
\Bigl\langle\mbox{in vac}\Bigl|
\sum_A\varepsilon_A\,
{\hat a}^{+}_{\mbox{\scriptsize out}}{}^A
{\hat a}_{\mbox{\scriptsize out}}{}^A
\Bigr|\mbox{in vac}\Bigr\rangle
\end{equation}
where $\varepsilon_A$ is the energy of the out-mode $A$, and
similarly for the other charges. This result needs no comments.
Miracles happen.
\subsubsection{Emission of Charges (Continued)}
An important point concerning miracles is that they happen
not always. Let us see what is needed for this particular 
miracle to happen. For that, it is necessary to introduce
characteristic parameters of the problem. There are two sets
of parameters.
\subparagraph{Parameters of the quantum field:}
$q$, $\mu$.
\subparagraph{Parameters of the external source:}
$e$, $l$, $\nu$.\par
\medskip
\noindent Here, $q$ and $\mu$ are the charge and mass of the
vacuum particles (e.g., of the electrons and positrons),
$e$ is the charge of the external source, $l$ is the 
characteristic width of its support tube, and $\nu$ is
the frequency parameter that characterizes the nonstationarity
of the source.

The vacuum current in (4.18) is of the form
\begin{equation}
J^{\mbox{\scriptsize vac}}=
\int\limits_0^\infty dm^2\,\rho(m^2)\frac{1}{m^2-
\Box_{\mbox{\scriptsize ret}}}
J^{\mbox{\scriptsize ext}}
+O\left(J^{\mbox{\scriptsize ext}}\right)^2\;.
\end{equation}
Here and above, the notation
$\Box_{\mbox{\scriptsize ret}}$
is to record that the resolvent is to be taken retarded.
The structure of the nonlinear terms in (4.30) is similar:
there is an overall resolvent acting on a function quadratic in
$J^{\mbox{\scriptsize ext}}$ (see (2.50)). If the vacuum particles
are massive, the spectral weight will be proportional to
the $\theta$-function:
\begin{equation}
\rho(m^2)\propto\theta(m^2-4\mu^2)
\end{equation}
to tell us that there is a threshold of pair creation. We need
to find the behaviour of
$J^{\mbox{\scriptsize vac}}$
at a large distance from the support of
$J^{\mbox{\scriptsize ext}}$:
\begin{equation}
J^{\mbox{\scriptsize vac}}\Bigl|_{r\gg l}=?
\end{equation}

First we need to calculate the action of the retarded resolvent
on a source $J^{\mbox{\scriptsize ext}}$ having a compact spatial
support. If $J^{\mbox{\scriptsize ext}}$ is static, the result is
\begin{equation}
\left.\frac{1}{m^2-\Box_{\mbox{\scriptsize ret}}}
J^{\mbox{\scriptsize ext}}\right|_{r\gg l}=
\frac{C}{r}\exp(-mr)\;,\;\quad
J^{\mbox{\scriptsize ext}}\mbox{ static.}
\end{equation}
At a large distance from the source, this is the Yukawa
potential. Because the function (4.33) is static, it does not
depend on the spacetime direction in which the limit
$r\gg l$ is taken. If
$J^{\mbox{\scriptsize ext}}$
is nonstatic, this is no more the case. The limit $r\gg l$
is direction-dependent, and there are directions in which
the decrease is slower. Namely, in the directions of the
outgoing light rays,
\begin{equation}
\left.\frac{1}{m^2-\Box_{\mbox{\scriptsize ret}}}
J^{\mbox{\scriptsize ext}}\right|_{r\gg l}=
\frac{C}{r}\exp\left(-m\sqrt{rU}\right)\;,\;\quad
J^{\mbox{\scriptsize ext}}\mbox{ nonstatic}
\end{equation}
where $U$ is a function of time\footnote{Of the retarded time
since the surfaces $\Sigma$ to which the outgoing light rays
belong are null.} whose order of magnitude is
\begin{equation}
U\sim\frac{1}{\nu}\;.
\end{equation}

Expression (4.34) is to be inserted in the spectral integral 
(4.30), and, since the spectrum is cut off from below, we find
that the vacuum current is suppressed by the factor
\begin{equation}
J^{\mbox{\scriptsize vac}}\sim\exp\left(
-\frac{\mu\sqrt{r}}{\sqrt{\nu}}\right)\;,\;\quad r\gg l\;.
\end{equation}
This is what constrains miracles. However, we find also that
the suppressing factor depends on the frequency of the source
and {\it can be removed by raising the frequency}. The farther
from the support of $J^{\mbox{\scriptsize ext}}$, the greater
the frequency should be for the current to be noticeable.
The pair creation starts as soon as the energy $\hbar\nu$
exceeds the threshold
\begin{equation}
\hbar\nu>2\mu c^2
\end{equation}
but, for the source to emit charge, the frequency should be
even greater:
\begin{equation}
\hbar\nu>(\mu c^2)\left(\frac{\mu c}{\hbar}l\right)\;.
\end{equation}
This is easy to understand. The particles start being created
in the support of the source with small momenta and cannot
go far away. The extra factor
$(\mu c/\hbar)l$
in (4.38) may be interpreted as the number of created particles
for which there is room in the support of the source. If the
creation is more violent, the particles get out of the tube.
This is the meaning of condition (4.38). The mechanism of emission
and conservation of charge is illustrated in Fig. 4. There are
initially the charges of the external source in its support tube.
They repel the like particles of the created pairs and,
when the number of the latter exceeds $(\mu c/\hbar)l$, push
them out of the tube. The unlike particles stay in the tube
and diminish its charge.
\input figfour.tex

Since the cause of the vacuum instability is the nonstationarity
of the external source, it is interesting to consider the case where
the energy $\hbar\nu$ exceeds overwhelmingly all the other energy
parameters of the problem. One can then study the strong effect
of particle production. It is assumed, in particular, that
$\hbar\nu$ exceeds both the rest energy of the vacuum particle
and its Coulomb energy in the external field:
\begin{equation}
\hbar\nu\gg\mu c^2\;,
\end{equation}
\begin{equation}
\hbar\nu\gg\frac{qe}{l}\;.
\end{equation}
In the limit (4.39), the flux of charge at a given distance
from the source ceases depending on the mass $\mu$, and
the vacuum particles can be considered as massless.
Condition (4.40) enables one to get rid of the consideration
of the static vacuum polarization which is irrelevant to
the problem. The approximation (4.39) and (4.40) is called
high-frequency approximation.

The effective action has been calculated above as an expansion
in powers of the curvature but the conditions of validity
of this expansion have not been discussed. This lack can now be met.
It is the high-frequency approximation in which this expansion
is valid. Indeed, consider the series (4.8). Every next term in
this series contains an extra power of ${\cal R}$, and, by dimension,
its formfactor contains an extra power of $\Box^{-1}$. The commutator
curvature is proportional to the charges and to $\hbar^{-1}$:
\begin{equation}
{\cal R}\sim\frac{qe}{\hbar l^2}\;.
\end{equation}
In the limit $r\gg l$ along the outgoing light rays, the
operator $\Box$ contains one time derivative:
\begin{equation}
\Box\sim\frac{\nu}{l}\;.
\end{equation}
As a result, every next term of the series contains, as
compared to the previous one, the extra factor
\begin{equation}
\frac{qe}{\hbar\nu l}\ll 1\;.
\end{equation}
In addition, the formfactors in (4.8) can be calculated
in the massless limit, as has been done above.

However, the inquest of miracles is not yet completed.
Assuming that the vacuum particles are massless or that
the high-frequency regime holds, we get rid of the suppressing
exponential in (4.36) but we still need to check the power
of decrease of the current. The power should be the one in
(4.23) for the emission of charge to occur. We can readily check this
since we know the behaviour of the resolvent. Expression (4.34)
is again to be inserted in the spectral integral (4.30) but this
time assuming that the spectrum begins with zero mass:
\begin{equation}
J^{\mbox{\scriptsize vac}}\Bigl|_{r\gg l}=
\int\limits_{\bf 0}^\infty dm^2\,\rho(m^2)
\frac{C}{r}\exp\left(-m\sqrt{rU}\right)\;.
\end{equation}
We see that, for the current to decrease as $O(1/r^2)$,
the spectral weight should have a finite and nonvanishing
limit at zero mass:
\begin{equation}
\rho(0)={}\mbox{finite}{}\ne 0\;.
\end{equation}
For the respective formfactor, this is a condition on its
behaviour at small $\Box$. The behaviour should be
\begin{equation}
F(\Box)=
\int\limits_0^\infty dm^2\,\frac{\rho(m^2)}{m^2-\Box}{}\;\quad
\parbox[t]{1cm}{
$\longrightarrow$\\$\Box\to 0$
}
{}-\rho(0)\ln (-\Box)\;.
\end{equation}

We arrive at the following consistency condition on the
vacuum formfactors. In the limit where one (any) of the
$\Box$ arguments is small and the others are fixed,
the formfactors should not grow faster than $\ln(-\Box)$:
\begin{equation}
F(\Box)\Bigl|_{\Box\to 0}=\mbox{const.}\ln(-\Box)\;,
\end{equation}
\begin{equation}
F(\Box_1,\Box_2,\Box_3)\Bigl|_{\Box_1\to 0}=
f(\Box_2,\Box_3)\ln(-\Box_1)\;,
\end{equation}
$$
\hbox to 190pt{{}\dotfill{}}
$$
If they grow faster, the charges cannot be maintained finite,
i.e., an isolated system cannot exist in such a vacuum.
If they grow as $\ln(-\Box)$, the theory of isolated systems
is consistent but these systems emit charges. If they grow
slower, the charges are conserved.

One can check whether the one-loop formfactors satisfy this
consistency condition. The second-order formfactors
(3.90)--(3.94) do. The third-order formfactors behave
generally as \cite{35}
\begin{equation}
F(\Box_1,\Box_2,\Box_3)\Bigl|_{\Box_1\to 0}=
f(\Box_2,\Box_3)\frac{1}{\Box_1}+
g(\Box_2,\Box_3)\ln(-\Box_1)+\ldots\;.
\end{equation}
The alarming terms $1/\Box$ appear only in the arguments
acting on the gravitational curvatures. Therefore, they
can affect only the vacuum energy-momentum tensor, and it
has been checked that, in the energy-momentum tensor,
these terms coming from different formfactors cancel.
In the currents, the one-loop formfactors satisfy strictly
the consistency condition. Since, in addition, their
asymptotic $\ln(-\Box)$ terms are nonvanishing, the emission
of charges in the high-frequency regime is real. The only
thing that remains to be checked is that this emission is
not a pure quantum noise. It will be checked by a direct
calculation.

Now one can answer also the question about the indefinite
local terms in the effective action. The coefficients of
these terms are the unspecified constants in (3.90)--(3.94).
In the limit $\Box\to 0$, the values of these constants are
immaterial. Only the terms $\ln(-\Box)$, $\Box\to 0$ of the
formfactors work, and, therefore, the incompleteness of
local quantum field theory does not affect the presently
considered problem.

It will be noted that there are now two mechanisms by which
an isolated system can emit energy. One is purely classical:
a nonstationary source can emit the electromagnetic or
gravitational waves. The other is quantum: 
immersed in the vacuum, 
a nonstationary source can emit also charged particles.
A high-frequency source will generally emit {\it both}.
\subsubsection{Particle Creation by External Fields}
The problem of particle creation by external fields is a part
of the expectation-value problem. In the context of the
foregoing, it can be set as follows. Consider the quantum
field that satisfies a linear second-order equation
\begin{equation}
\left(g^{\mu\nu}\nabla_\mu\nabla_\nu{\hat 1}+{\hat P}\right)
\phi=0
\end{equation}
containing three external fields: the metric, the connection,
and the potential. The external fields are asymptotically
static in the past and future but otherwise arbitrary except
that their currents
\begin{eqnarray}
J_{\alpha\beta}&=&R_{\alpha\beta}-\frac{1}{2}g_{\alpha\beta}R\;,\\
{\hat J}_\alpha&=&\nabla^\beta{\hat{\cal R}}_{\alpha\beta}\;,\\
{\hat Q}&=&{\hat P}+\frac{1}{6}R{\hat 1}
\end{eqnarray}
are confined to a spacetime tube. The quantum field is in the
in-vacuum state. What is the energy of the quanta of the field
$\phi$ created by the external fields for the entire history?
In the high-frequency approximation, we have everything to
answer this question.

To formulate the answer, I need some preliminary construction.
Every current has an associated quantity called its radiation
moment. It will now be defined.

Consider a timelike geodesic in the external metric of
equation (4.50). It enters the domain of nonstationarity of external
fields with a definite energy and goes out of this domain with
a definite energy. Let $E$ be its energy per unit rest mass
on going out. I am only interested in the geodesics that
escape to $r=\infty$. They have $E>1$, and, instead of $E$,
I shall use the parameter $\gamma$ defined as
\begin{equation}
\gamma=\frac{\sqrt{E^2-1}}{E}\;,\quad E>1\;,\quad
0<\gamma<1\;.
\end{equation}
At $r=\infty$, the geodesic has a certain spatial direction, or,
equivalently, it comes to a certain point of the celestial 
2-sphere. I shall denote this sphere as ${\cal S}$, its points
as $\theta$:
\begin{equation}
\theta=(\theta_1,\theta_2)\;,\;\quad\theta\in{\cal S}\;,
\end{equation}
and the integral over the unit 2-sphere as
\begin{equation}
\int d^2{\cal S}(\theta)\,(\cdots)\;.
\end{equation}
A geodesic with given $\gamma$ and $\theta$ will be called
$\gamma,\theta$ -geodesic (see Fig. 5).
\input figfive.tex

A $\gamma,\theta$ -geodesic can be emitted from every point
of a compact domain. Therefore, the $\gamma,\theta$ -geodesics
with {\it the same values} of $\gamma$ and $\theta$ make a
congruence, and it can be proven that this congruence is
hypersurface-orthogonal. Let the orthogonal hypersurfaces be
\begin{equation}
T_{\gamma\theta}(x)=\mbox{const.}
\end{equation}
Since the parameters $\gamma,\theta$ fix the congruence,
they fix also the family of the orthogonal hypersurfaces (4.57),
and the "const." in (4.57) fixes a member of the family. The
function $T_{\gamma\theta}$ is determined up to a transformation
$T_{\gamma\theta}\to f\left(T_{\gamma\theta}\right)$. This
arbitrariness will be removed by the normalization condition
\begin{equation}
\left(\nabla T_{\gamma\theta}\right)^2=-\left(1-\gamma^2\right)
\end{equation}
and the condition that the vector $\nabla T_{\gamma\theta}$
is past directed. It is a property of the geodetic congruences
that the norm in (4.58) can be chosen constant.

The radiation moment of any scalar current $J$ is the following
hypersurface integral:
\begin{equation}
D=\frac{1}{4\pi}\int dx\,g^{1/2}\delta
\left(T_{\gamma\theta}(x)-\tau\right)J(x)\;.
\end{equation}
If the current is not a scalar, it should first be parallel
transported from the integration point to $r=\infty$ along
the respective $\gamma,\theta$ -geodesic. Thus if the current
is a vector, its radiation moment is
\begin{equation}
D^\alpha=\frac{1}{4\pi}\int dx\,g^{1/2}\delta
\left(T_{\gamma\theta}(x)-\tau\right)J^\beta(x)
a_\beta{}^\alpha(x,\infty)
\end{equation}
where $a_\beta{}^\alpha(x,\infty)$ is the propagator of
parallel transport of vectors to infinity along the
$\gamma,\theta$ -geodesic emanating from $x$. The radiation
moment $D^\alpha$ is then a vector at infinity. In the same
way, the radiation moment is defined for any current.
For the three currents (4.51)--(4.53), the radiation moments will
be denoted respectively as
\begin{equation}
J_{\alpha\beta},\;{\hat J}_\alpha,\;{\hat Q}
\longrightarrow D_{\alpha\beta},\;{\hat D}_\alpha,\;{\hat D}\;.
\end{equation}
Since the indices of the radiation moments pertain to a point
at infinity, their contractions like
\begin{equation}
{\hat D}_\alpha{\hat D}^\alpha=g_{\alpha\beta}
{\hat D}^\alpha{\hat D}^\beta\;,\quad\mbox{etc.}
\end{equation}
always assume the flat metric $g_{\alpha\beta}$ at infinity.
All radiation moments are functions of four parameters:
\begin{equation}
D=D(\gamma,\theta,\tau)\;.
\end{equation}

In the limit $\gamma=1$, the $\gamma,\theta$ -geodesics become null.
The orthogonal hypersurfaces (4.57) also become null, and the geodesics
themselves become their generators. For the radiation moments,
this is a regular limit. Nothing special happens to them in this
limit except that they become very important. The radiation
moments at $\gamma=1$ govern the emission of waves in classical
theory. Thus if $J_\alpha$ in (4.52) is an electric current,
then the following expression:
\begin{eqnarray}
&&\Bigl(M(-\infty)-M(+\infty)\Bigr)_{\mbox{electromagnetic waves}}
\nonumber\\
&&\hphantom{=}
{}=\frac{1}{4\pi}\int\limits_{-\infty}^\infty d\tau
\int d^2{\cal S}(\theta)\,
\left.\left[g_{\alpha\beta}
\left(\frac{d}{d\tau}D^\alpha\right)
\left(\frac{d}{d\tau}D^\beta\right)\right]\right|_{\gamma=1}
\qquad\qquad\qquad\qquad\;
\end{eqnarray}
is the energy of the electromagnetic waves emitted by this
current for the entire history. A similar expression with
the tensor current (4.51):
\begin{eqnarray}
&&\Bigl(M(-\infty)-M(+\infty)\Bigr)_{\mbox{gravitational waves}}
\nonumber\\
&&{}=\frac{1}{4\pi}\int\limits_{-\infty}^\infty d\tau
\int d^2{\cal S}(\theta)\,
\frac{1}{2}(g_{\alpha\mu}g_{\beta\nu}-
\frac{1}{2}g_{\alpha\beta}g_{\mu\nu})
\left(\frac{d}{d\tau}D^{\alpha\beta}\right)
\left.\left(\frac{d}{d\tau}D^{\mu\nu}\right)\right|_{\gamma=1}
\quad\;\nonumber\\
\end{eqnarray}
is the energy of the gravitational waves emitted by the
current $J_{\alpha\beta}$ for the entire history.

The radiation moment is a generating function for the
multipole moments. The multipole expansion is the
expansion of $D$ at $\gamma=0$. It makes sense for
nonrelativistic systems since $\gamma$ is proportional
to $1/c$.

Expressions (4.64) and (4.65) are the solutions of the classical
radiation problem. And here is the solution of the
quantum radiation problem \cite{50}:
\begin{eqnarray}
&&\Bigl(M(-\infty)-M(+\infty)\Bigr)_{\mbox{created particles}}
\nonumber\\
&&\hphantom{=}
{}=\frac{1}{(4\pi)^2}
\int\limits_0^1 d\gamma\,\gamma^2
\int\limits_{-\infty}^\infty d\tau
\int d^2{\cal S}(\theta)\mathop{\rm tr}\left[
\left(\frac{d^2}{d\tau^2}{\hat D}\right)^2\right.\nonumber\\
&&\hphantom{={}=}\qquad\qquad\qquad\quad
{}-\frac{1}{3}\frac{1}{(1-\gamma^2)}g_{\alpha\beta}
\left(\frac{d}{d\tau}{\hat D}^\alpha\right)
\left(\frac{d}{d\tau}{\hat D}^\beta\right)\nonumber\\
&&\hphantom{={}=}\qquad\qquad\qquad\quad
{}+\frac{1}{30}{\hat 1}
(g_{\alpha\mu}g_{\beta\nu}-
\frac{1}{3}g_{\alpha\beta}g_{\mu\nu})
\left.\left(\frac{d^2}{d\tau^2}D^{\alpha\beta}\right)
\left(\frac{d^2}{d\tau^2}D^{\mu\nu}\right)\right]\;.
\nonumber\\
\end{eqnarray}
This is the energy of the quanta of the field $\phi$ created
by the external fields for the entire history. As compared
to the expressions above, there is an extra time 
derivative in the case of the tensor and scalar moments. It accounts
for the dimension of the coupling constant. Also, instead
of setting $\gamma=1$, one needs to integrate over $\gamma$.
Otherwise, the similarity is striking. The quantum problem
of particle creation becomes almost the same thing as the
classical problem of emission of waves.

The presence in (4.66) of an integral over $\gamma$ is not just
a technical detail. The radiation moments have both the
longitudinal projections, i.e., the projections on the
direction of the geodesic at infinity and the transverse
projections. Inspecting the contractions of the moments
in (4.64)--(4.66), one can see that, at $\gamma=1$, the 
longitudinal projections drop out of these contractions.
In the integral over $\gamma$, also the longitudinal
projections survive. Owing to this fact, spherically
symmetric sources cannot emit waves but can produce
particles from the vacuum.

Now I can explain why, when expanding the effective action,
I stopped at the terms cubic in the curvature. In the
high-frequency approximation, the expansion (3.40) needs
to be calculated up to the lowest-order terms that give
a nonvanishing effect. The terms of first order in the
curvature are local and give no effect. The terms of
second order in the curvature are nonlocal and contribute
to the energy flux at infinity but it turns out that
{\it their contribution is a pure quantum noise}. The
real effect of particle production begins with the third
order in the curvature. Expression (4.66) results from
the triangular loop diagrams.

Since varying the action destroys one curvature, a cubic 
action generates a quadratic current. This gives the radiation
energy a chance to be positive definite. Expression (4.66)
is positive definite indeed:
\begin{equation}
\Bigl(M(-\infty)-M(+\infty)\Bigr)_{\mbox{created particles}}
\ge 0\;.
\end{equation}
In particular, for the matrix contributions, this follows
from relations (3.11), (3.12) and the positive definiteness
of the matrix $\omega_{ab}$:
\begin{equation}
\mathop{\rm tr}
\left(\frac{d^2}{d\tau^2}{\hat D}\right)^2\ge 0\;,\;\quad
\mathop{\rm tr}\left[g_{\alpha\beta}
\left(\frac{d}{d\tau}{\hat D}^\alpha\right)
\left(\frac{d}{d\tau}{\hat D}^\beta\right)\right]\le 0\;.
\end{equation}
The positivity of the gravitational-field contribution can
be proven directly.
\subsubsection{The Backreaction Problem}
The energy emitted by an isolated system (in all forms) should
be bounded both from below and from above: it should be positive
and less than the energy stored in the initial state
\begin{equation}
0\le\Bigl(M(-\infty)-M(+\infty)\Bigr)\le M(-\infty)\;.
\end{equation}
In expression (4.66), the positivity is guaranteed but the
energy conservation is not. The reason is that the setting 
of the problem with external fields is physically inconsistent.
The vacuum current determines the solution of the mean-field
equations, and the mean field rather than the external field
determines the vacuum current. If the backreaction of the vacuum
is neglected, the conservation laws need not be observed.

One case in which the vacuum backreaction may not be neglected
is where both mechanisms of the energy emission, classical
and quantum, are engaged simultaneously. This concerns
particularly the vector connection field. In expression (4.66),
the integral over $\gamma$ has a pole ${(1-\gamma)^{-1}}$ in
the term with the vector moment. The residue of the integrand
in this pole is precisely the quantity (4.64), i.e., the energy
of the outgoing waves of the vector connection field. If it is
nonvanishing, e.g., if the external source emits both the
electromagnetic waves and the electrically charged particles, the
integral in $\gamma$ diverges. The result is a disaster: the
radiation energy appears to be infinite. In fact it should be
taken into account that the created charge affects the
generation of the electromagnetic waves, and the respective
changes in the electromagnetic field affect the creation of charge.
In the self-consistent solution, the disaster is removed.

Another example concerns the metric field when it has an event
horizon. In this case, the integral in $\tau$ diverges at the
upper limit. By construction, $\tau$ is the time of an external
observer. As $\tau\to\infty$, the source moving in the tube
hits the event horizon. Its proper time does not turn into
infinity. The integrand in (4.66) is just finite in this limit,
and the integral in $\tau$ diverges linearly. This is the Hawking
constant flux of radiation from the black hole. If its
backreaction on the metric is neglected, the total emitted
energy is infinite.

But even when the quantity (4.66) is finite, it depends on the
frequency of the source. If the source is external, this
frequency is a free parameter. The energy of created quanta
grows with frequency, and, typically, the ratio
\begin{equation}
\left.\frac{M(-\infty)-M(+\infty)}{M(-\infty)}\right|_{\nu\to\infty}
\sim\ln\nu
\end{equation}
also grows so that, at a sufficiently high frequency, the energy
conservation law will be violated. The backreaction should take
into account that, when the source creates real particles,
it loses energy and slows down. It then creates less particles,
and the process dies away. The conservation laws will then be restored.

The backreaction problem has been solved only in a few cases
\cite{51}--\cite{56}. The examples for which it has been solved
show that the
solution can be unexpected and interesting.
}

%% file: figthree.tex
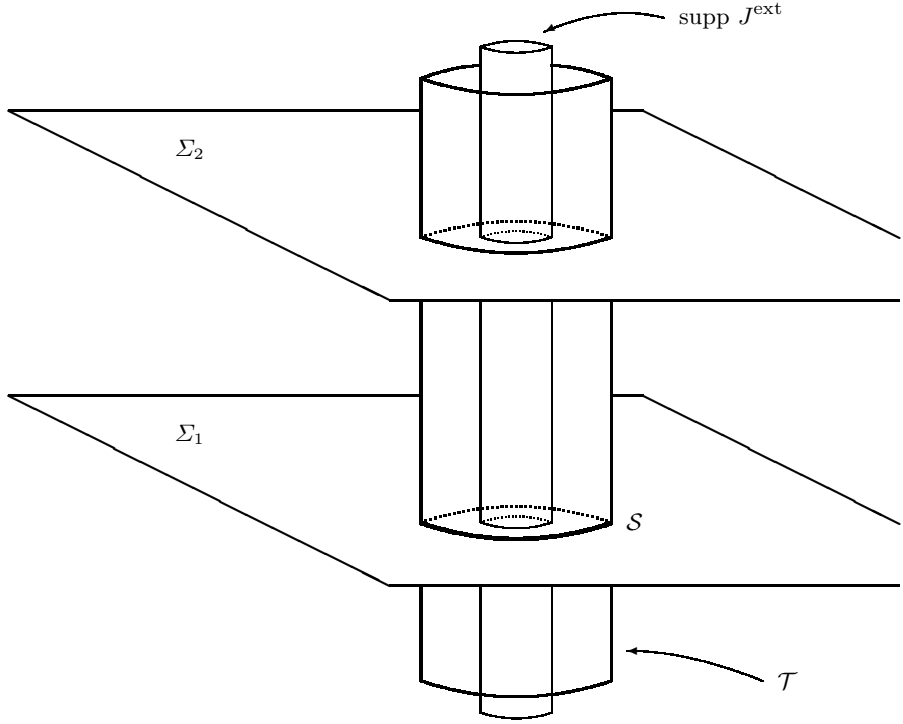
\begin{figure}
\centering
\begin{picture}(337,288.1)(0,-60)
\thicklines
\put(0,72){\line(2,-1){144}}
\put(240,72){\line(2,-1){97}}
\put(0,180){\line(2,-1){144}}
\put(240,180){\line(2,-1){97}}
\put(0,72){\line(1,0){156}}
\put(228,72){\line(1,0){12}}
\put(0,180){\line(1,0){156}}
\put(228,180){\line(1,0){12}}
\put(144,0){\line(1,0){193}}
\put(144,108){\line(1,0){193}}
\put(156,192){\line(0,-1){60}}
\put(228,192){\line(0,-1){60}}
\put(156,108){\line(0,-1){84}}
\put(228,108){\line(0,-1){84}}
\put(156,0){\line(0,-1){36}}
\put(228,0){\line(0,-1){36}}
\qbezier(156,-36)(192,-48)(228,-36)
\qbezier(156,24)(192,12)(228,24)
\qbezier(156,23.5)(192,11.5)(228,23.5)
\qbezier[40](156,24)(192,36)(228,24)
\qbezier(156,132)(192,120)(228,132)
\qbezier[40](156,132)(192,144)(228,132)
\qbezier(156,192)(192,180)(228,192)
\qbezier(156,192)(167.4,196.5)(178.5,197)
\qbezier(228,192)(216.6,196.5)(205.5,197)
{\thinlines
\put(178.5,204){\line(0,-1){72}}
\put(205.5,204){\line(0,-1){72}}
\put(178.5,108){\line(0,-1){84}}
\put(205.5,108){\line(0,-1){84}}
\put(178.5,0){\line(0,-1){48}}
\put(205.5,0){\line(0,-1){48}}
\qbezier(178.5,-48)(192,-52.5)(205.5,-48)
\qbezier(178.5,24)(192,19.5)(205.5,24)
\qbezier[20](178.5,24)(192,28.5)(205.5,24)
\qbezier(178.5,132)(192,127.5)(205.5,132)
\qbezier[20](178.5,132)(192,136.5)(205.5,132)
\qbezier(178.5,204)(192,199.5)(205.5,204)
\qbezier(178.5,204)(192,208.5)(205.5,204)
\qbezier(248,214.86)(228,220.57)(202.64,209.71)
\put(205.64,211.21){\vector(-2,-1){3}}
\put(253.71,212.717){supp $J^{\mbox{\scriptsize ext}}$}
\qbezier(285.14,-36)(256.57,-24.57)(233.71,-24.57)
\put(236.71,-24.57){\vector(-1,0){3}}
\put(290.85,-38.86){${\cal T}$}
\put(62.9,162.76){$\Sigma_2$}
\put(62.9,54.76){$\Sigma_1$}
\put(233.71,21.14){${\bf\cal S}$}
}
\end{picture}
\caption{Support tube of $J^{\mbox{\scriptsize ext}}$ and
a wider tube.}
\label{fig:3}
\end{figure}

%% file: figfour.tex
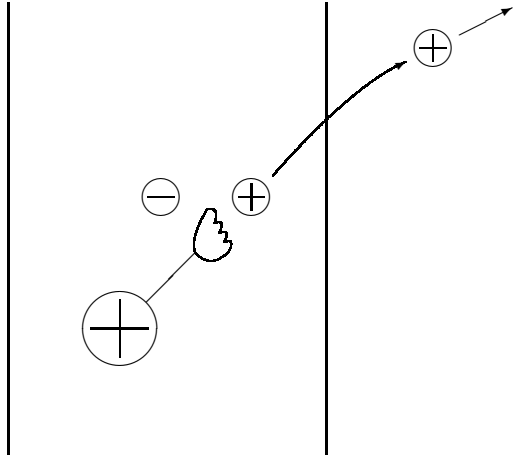
\begin{figure}
\centering
\begin{picture}(120.37,171)(-40,-45.71)
\put(2.02,2.02){\begin{picture}(66.085,66.085)%
\put(0,0){\circle{28.57}}
\put(0,0){\line(1,0){11}}
\put(0,0){\line(-1,0){11}}
\put(0,0){\line(0,1){11}}
\put(0,0){\line(0,-1){11}}
\put(10.1,10.1){\line(1,1){18.47}}
\qbezier(28.57,28.57)(26.55,34.62)(33.116,45.226)
\qbezier(28.57,28.57)(34.62,22.51)(41.196,29.066)
\qbezier(33.116,45.226)(37.997,46.067)(35.81,39.84)
\qbezier(35.81,39.84)(40.242,41.578)(37.603,36.253)
\qbezier(37.603,36.253)(42.038,37.986)(39.403,32.653)
\qbezier(39.403,32.653)(43.833,34.393)(41.196,29.066)
\end{picture}}
\put(51.8,51.8){\circle{14.28}}
\put(17.51,51.8){\circle{14.28}}
\put(51.8,51.8){\line(1,0){5}}
\put(51.8,51.8){\line(-1,0){5}}
\put(51.8,51.8){\line(0,1){5}}
\put(51.8,51.8){\line(0,-1){5}}
\put(17.51,51.8){\line(1,0){5}}
\put(17.51,51.8){\line(-1,0){5}}
{\thicklines
\put(-40,-45.71){\line(0,1){171}}
\put(80.37,-45.71){\line(0,1){171}}
}
%
\put(120.37,108.15){\circle{14.28}}
\put(120.37,108.15){\line(1,0){5}}
\put(120.37,108.15){\line(-1,0){5}}
\put(120.37,108.15){\line(0,1){5}}
\put(120.37,108.15){\line(0,-1){5}}
\qbezier(59.88,59.88)(90.37,94.37)(110.15,103.04)
\put(107.466,101.698){\vector(2,1){3}}
\put(130.59,113.26){\vector(2,1){20}}
\end{picture}
\caption{Mechanism of emission and conservation of charge.}
\label{fig:4}
\end{figure}

%% file: figfive.tex
\begin{figure}
\centering
\begin{picture}(300,202)(0,-56)
{\thicklines
\put(0,0){\line(1,0){300}}
\put(0,60){\line(1,0){300}}
}
\multiput(0,0)(10,0){25}{\line(1,1){60}}
\multiput(0,60)(10,0){25}{\line(1,-1){60}}
\put(250,0){\line(1,1){50}}
\put(260,0){\line(1,1){40}}
\put(270,0){\line(1,1){30}}
\put(280,0){\line(1,1){20}}
\put(290,0){\line(1,1){10}}
\put(0,10){\line(1,1){50}}
\put(0,20){\line(1,1){40}}
\put(0,30){\line(1,1){30}}
\put(0,40){\line(1,1){20}}
\put(0,50){\line(1,1){10}}
\put(0,50){\line(1,-1){50}}
\put(0,40){\line(1,-1){40}}
\put(0,30){\line(1,-1){30}}
\put(0,20){\line(1,-1){20}}
\put(0,10){\line(1,-1){10}}
\put(290,60){\line(1,-1){10}}
\put(280,60){\line(1,-1){20}}
\put(270,60){\line(1,-1){30}}
\put(260,60){\line(1,-1){40}}
\put(250,60){\line(1,-1){50}}
{\thicklines
\put(122,60){\vector(1,2){43}}
\put(92,0){\line(-1,-2){28}}
\put(67.5,27.15){{\bf DOMAIN OF NONSTATIONARITY}}
}
\put(169.6,143){$r\to\infty$}
\put(169.6,131){parameters: $\gamma,\theta$}
\end{picture}
\caption{A $\gamma,\theta$ -geodesic.}
\label{fig:5}
\end{figure}
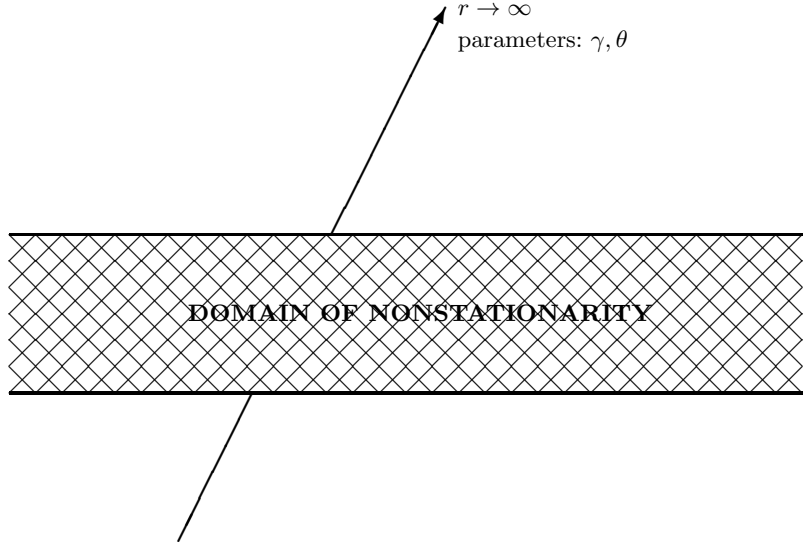